\documentclass[a4paper,fleqn,usenatbib]{mnras}

\usepackage{mathptmx}

\usepackage[T1]{fontenc}
\usepackage{ae,aecompl}

\usepackage{graphicx}	
\usepackage{amsmath}	
\usepackage{amssymb}	
\usepackage{bm}
\usepackage[dvipsnames]{xcolor}

\newcommand{\vektor}[1]{\bm{#1}}

\newcommand{\percc}{\mathrm{cm}^{-3}}
\newcommand{\gpercc}{\mathrm{g\,cm}^{-3}}
\newcommand{\kmpers}{\mathrm{km\,s}^{-1}}
\newcommand{\Bparv}{\vektor{B}\parallel\vektor{v}_\mathrm{w}}
\newcommand{\Bperpv}{\vektor{B}\perp\vektor{v}_\mathrm{w}}

\title[In situ H$_2$ formation in a hot outflow]{The in situ formation of molecular and warm ionised gas triggered by hot galactic outflows}

\author[Philipp Girichidis et al.]{
  Philipp~Girichidis$^{1}$\thanks{E-mail: philipp@girichidis.com},
  Thorsten~Naab$^{2}$,
  Stefanie~Walch$^{3}$,
  Thomas~Berlok$^{1}$
\\
$^1$Leibniz-Institut f\"{u}r Astrophysik (AIP), An der Sternwarte 16, 14482 Potsdam, Germany\\
$^2$Max-Planck-Institut f\"{u}r Astrophysik, Karl-Schwarzschild-Str. 1, 85741 Garching, Germany\\
$^3$I. Physikalisches Institut, Universit\"{a}t zu K\"{o}ln, Z\"{u}lpicher Str. 77, 50937 K\"{o}ln, Germany
}

\date{Accepted XXX. Received YYY; in original form ZZZ}

\pubyear{2021}

\begin{document}
\label{firstpage}
\pagerange{\pageref{firstpage}--\pageref{lastpage}}
\maketitle

\begin{abstract}
Molecular outflows contributing to the matter cycle of star forming galaxies are now observed in small and large systems at low and high redshift. Their physical origin is still unclear. In most theoretical studies only warm ionised/neutral and hot gas outflowing from the interstellar medium is generated by star formation. We investigate an in situ H$_2$ formation scenario in the outflow using high-resolution simulations, including non-equilibrium chemistry and self-gravity, of turbulent, warm, and atomic clouds with densities 0.1, 0.5 and $1\,\percc$ exposed to a magnetised hot wind. For cloud densities $\gtrsim 0.5\,\percc$ a magnetised wind triggers H$_2$ formation before cloud dispersal. Up to 3 per cent of the initial cloud mass can become molecular on $\sim 10\,\mathrm{Myr}$ time scales. The effect is stronger for winds with perpendicular $B$-fields and intermediate density clouds ($n_\mathrm{c}\sim 0.5\,\percc$). Here H$_2$ formation can be boosted by up to one order of magnitude compared to isolated cooling clouds independent of self-gravity. Self-gravity preserves the densest clouds well past their $\sim 15\,\mathrm{Myr}$ cloud crushing time scales. This model could provide a plausible in situ origin for the observed molecular gas. All simulations form warm ionised gas, which represents an important observable phase. The amount of warm ionised gas is almost independent of the cloud density but solely depends on the magnetic field configuration in the wind. For low density clouds ($0.1\,\percc$), up to 60 per cent of the initially atomic cloud mass can become warm and ionised.
\end{abstract}

\begin{keywords}
ISM: clouds -- ISM: magnetic fields -- ISM: abundances -- galaxies: ISM -- magnetohydrodynamics (MHD) -- instabilities 
\end{keywords}



\section{Introduction}

The matter cycle in galaxies is tightly coupled to the thermal and dynamical processes in the interstellar medium, in particular heating and cooling \citep[e.g.][]{McKeeOstriker1977, Draine2011, KlessenGlover2016}, the formation of dense gas that forms stars \citep[e.g.][]{GirichidisEtAl2020b} as well as galactic disc inflows and outflows \citep[e.g.][]{NaabOstriker2017}. Here, molecular hydrogen (H$_2$) and carbon monoxide (CO) represent the dense and often self-gravitating phase which mostly correlates with the formation of stars \citep[e.g.][]{BigielEtAl2008}. Cold and molecular gas is also observed in galactic outflows \citep[e.g.][]{VeilleuxEtAl2020, FoersterSchreiberStijn2020}. There is growing evidence for molecular outflows associated with star formation in galaxies. This not only applies to the well-known case of M82 \citep{WalterWeissScoville2002, LeroyEtAl2015} but also to other nearby star forming disc galaxies \citep[e.g.][]{KriegerEtAl2019, LutzEtAl2020, SalakEtAl2020}, the Milky Way \citep{DiTeodoroEtAl2020} and even dwarf galaxies like the SMC \citep{DiTeodoroEtAl2019}. Molecular outflows associated with star formation have also been detected up to redshift $z\sim4$ \citep[e.g.][]{SpilkerEtAl2020a,SpilkerEtAl2020b}. However, the process of how molecular gas (H$_2$ and CO) ends up in outflows is not yet understood. It could be accelerated from the midplane or form within a dense outflow -- or a combination of both. For starburst environments analytical models of cooling shells provide theoretical predictions in agreement with observations \citep{RoyEtAl2016}.

The formation of dense and molecular gas in the disc has been modelled successfully for individual clouds \citep[e.g.][]{GloverClark2012a, ValdiviaEtAl2016} as well as for an ensemble of clouds in the multiphase interstellar medium \citep[e.g.][]{WalchEtAl2015, SeifriedEtAl2017, SeifriedEtAl2020Zoom, BellomiEtAl2020, SmithREtAl2020}. Simulations of the multiphase interstellar medium are able to model feedback-driven outflows using supernovae \citep[e.g.][]{GirichidisEtAl2016b, MartizziEtAl2016, LiBryanOstriker2017, FieldingQuataertMartizzi2018, KimOstriker2018}, stellar winds \citep[e.g.][]{GattoEtAl2017} and radiation \citep[e.g.][]{PetersEtAl2017a, RathjenEtAl2021} but only hot and warm gas is accelerated in models of the ISM. Similarly, outflows driven in dwarf galaxies are not cold \citep[e.g.][]{HuEtAl2016, EmerickBryanMacLow2019, Hu2019, DashanDubois2020, SmithMEtAl2020, GutckeEtAl2020, SchneiderEtAl2020}. Even cosmic ray-driven outflows, which are typically denser and cooler do not contain measurable contents of cold or molecular gas \citep[e.g.][]{GirichidisEtAl2016a, SimpsonEtAl2016, GirichidisEtAl2018a, SemenovKravtsovCaprioli2020, RathjenEtAl2021}. Both the formation of molecular gas in the outflow as well as the acceleration of molecular gas from the disc pose challenges to the models in particular because of limited resolution in galactic scale simulations.

Complementary to global models are idealised wind tunnel simulations focussing on the evolution of a local overdensity exposed to a dilute hot wind \citep[e.g.][]{KleinMcKeeColella1994, PittardEtAl2005, VieserHensler2007, BrueggenScannapieco2016, GronkeOh2018}. Besides the acceleration of dense gas, the survival of clouds and their potential growth due to condensing hot gas are of particular interest \citep{GronkeOh2020a}. The mixing of hot ionized gas with warm gas can form warm ionized gas (WIM), which is the most commonly used phase to study outflows \citep[e.g.][]{FoersterSchreiberEtAl2019}. Radiative cooling plays a pivotal role for the survival of clouds. However, most numerical models to date focus on the hot and warm gas with temperatures down to $\sim10^4\,\mathrm{K}$. The cloud evolution including the cold and molecular gas phase down to star-forming temperatures have only be recently assessed \citep{BandaBarraganEtAl2021} using simplified tabulated cooling tables or are studied on much smaller spatial scales in the context of the onset of gravitational collapse \citep{JohanssonZiegler2013}. Models that include relevant non-equilibrium chemical evolution in the dense gas below $10^4\,\mathrm{K}$ are still missing.  

Magnetic fields are an essential part of galaxies \citep[e.g.][]{Crutcher2012, Haverkorn2015} with energy densities comparable to those in the other components of the ISM. Strong fields influence the gas dynamics and can redirect the flow, where the orientation of the field plays a critical role. In ideal magneto-hydrodynamics (MHD), which is a good approximation for most of the ISM, the field lines are frozen in the gas \citep[e.g.][]{Spruit2013}, such that gas flowing perpendicular to the field drags the field lines with it. This means in turn, that a magnetised flow with a field orientation perpendicular to the direction of the stream can efficiently transport the gas and accelerate it. This fact leads to magnetic draping, in which the field lines are bent around a local over-density.

In this study we would like to address the problem of the condensation of molecular hydrogen out of the warm neutral medium in a hot and dynamic environment. We perform idealised simulations of a wind tunnel experiment, in which a cloud of neutral gas is exposed to a hot ionised wind. We investigate how the orientation of the magnetic field influences the evolution of the clouds, the cooling of gas and the formation of molecular hydrogen.

The paper is organized as follows. In section~\ref{sec:numerics} we discuss the numerical setup, the physical processes included as well as the numerical parameters. Section~\ref{sec:char-numbers} presents characteristic numbers of clouds in a wind. The presentation of the results starts with the morphological evolution in Section~\ref{sec:morphology}, followed by the effective acceleration that the cloud experiences (Section~\ref{sec:acceleration}). Section~\ref{sec:turbulence} focusses on the turbulence that the wind drives into the cloud. In Section~\ref{sec:chemistry} we discuss the formation of dense gas and molecular hydrogen. The discussion of the results and the conclusions are presented in Section~\ref{sec:discussion-conclusion}.

\section{Numerical Methods and Simulation parameters}
\label{sec:numerics}

\subsection{Numerical code}

We use the hydrodynamical code FLASH \citep{FLASH00, DubeyEtAl2008} in version 4. The MPI-parallelized code uses adaptive mesh refinement (AMR) based on the PARAMESH library \citep{PARAMESH99}. PARAMESH splits the domain into blocks of size $8^3$. The ideal magneto-hydrodynamic equations are solved using the HLLR3 solver \citep{Bouchut2007, Bouchut2010,Waagan2009,Waagan2011}, which ensures positivity of the gas density and the pressure and is suitable for highly dynamical systems and high Mach number flows. We solve the following set of equations
\begin{align*}
\frac{\partial\rho}{\partial t} + \vektor{\nabla}\cdot\left(\rho\vektor{v}\right) &= 0\\
\frac{\partial\rho\vektor{v}}{\partial t} + \vektor{\nabla}\cdot\left[\rho\vektor{v}\vektor{v}^\mathrm{T}+\left(P_\mathrm{th} + \frac{\vektor{B}^2}{8\pi}\right)\mathsf{I} - \frac{\vektor{B}\vektor{B}^\mathrm{T}}{4\pi}\right] &= \rho\vektor{g}\\
\frac{\partial e}{\partial t} + \vektor{\nabla}\cdot\left[\left(u + \frac{\rho\vektor{v}^2}{2} + \frac{\vektor{B}^2}{8\pi} + \frac{P_\mathrm{th}}{\rho}\right)\vektor{v} - \frac{\vektor{B}\left(\vektor{v}\cdot\vektor{B}\right)}{4\pi}\right] &= \rho \vektor{v}\cdot\vektor{g} +\dot{u}_\mathrm{chem}\\
\frac{\partial\vektor{B}}{\partial t} - \vektor{\nabla}\times\left(\vektor{v}\times\vektor{B}\right) &= 0,
\end{align*}
with gas density $\rho$ and velocity $\vektor{v}$. The magnetic field, $\vektor{B}$, is described in the Gaussian unit system. The thermal energy density is given by $u$, the thermal pressure by $P_\mathrm{th}$ and the total energy by
\begin{equation}
e = u + \frac{\rho\vektor{v}^2}{2} + \frac{\vektor{B}^2}{8\pi}.
\end{equation}
The notation $\vektor{x}\vektor{x}^\mathrm{T}$ denotes the dyadic product of two vectors. We close the system with the equation of state
\begin{equation}
P_\mathrm{th} = (\gamma-1)\,u.
\end{equation}
The gravitational acceleration, $\vektor{g} = -\nabla\Phi$, due to self-gravity is computed by solving the Poisson equation
\begin{equation}
\vektor{\nabla}^2 \Phi = 4\pi G \rho
\end{equation}
with the gravitational potential $\Phi$ and the gravitational constant $G$ using the tree-based solver by \citet{WuenschEtAl2018} with the standard settings. Thermal heating and cooling is denoted by $\dot{u}_\mathrm{chem}$ and includes energy changes due to chemical transitions, radiative cooling as well as background UV and X-ray heating. The changes in thermal energy are included via a chemical network, which includes the non-equilibrium abundances of ionized (H$^+$), atomic (H) and molecular (H$_2$) hydrogen, carbon monoxide (CO), singly ionized carbon (C$^+$), and free electrons. The implementation in FLASH and the chosen parameters of the chemical model are identical to those in the SILCC simulations \citep{WalchEtAl2015, GirichidisEtAl2016b}. Here, we only briefly summarize the modules. The details of the chemical network are described in \citet{GloverMacLow2007a} and \citet{MicicEtAl2012}, the carbon chemistry in the network follows the model by \citet{NelsonLanger1997}. The atomic and molecular cooling functions assume solar metallicity and follow the models by \citet{GloverEtAl2010} and \citet{GloverClark2012b}. At temperatures above $10^4\,\mathrm{K}$ we assume collisional ionisation equilibrium and apply the rates from \citet{GnatFerland2012}. We use total carbon, oxygen and silicon abundances of $x_\mathrm{C,tot}=1.41\times10^{-4}$, $x_\mathrm{O,tot}=3.16\times10^{-4}$, $x_\mathrm{Si,tot}=1.5\times10^{-5}$, respectively \citep{SembachEtAl2000}. We assume H$_2$ to form on the surface of dust grains \citep{Hollenbach89} with a constant dust-to-gas mass ratio of 0.01 independent of the temperature, i.e. we do not form or destroy dust. The dust temperature is computed self-consistently, see \citet{WalchEtAl2015} for details.

We assume a temporally constant background UV heating with a strength of $1.7\,G_0$ where $G_0$ is the integrated energy density from $6$ to $13.6\,\mathrm{eV}$ in units of the Habing estimate $u_\mathrm{Habing}= 5.29 \times 10^{-14}\,\mathrm{erg\,cm}^{-3}$ \citep{Habing1968, Draine1978, Draine2011}. The radiation is locally attenuated in optically thick regions. We compute the shielding of individual cells using the TreeCol algorithm \citep{ClarkGloverKlessen2012} with the attenuation factors described in \citet{GloverClark2012b}. The implementation is described in \citet{WuenschEtAl2018} as the TreeRay/OpticalDepth module.

We adopt a cosmic ray ionization rate of $\zeta_\mathrm{CR} = 3 \times 10^{-17}\,\mathrm{s}^{-1}$ which corresponds to a heating term as described in \citet{GoldsmithLanger1978}. In addition, X-ray heating \citep{WolfireEtAl1995} and photoelectric heating are included and coupled to the optical depth \citep{BakesTielens1994, WolfireEtAl2003, Bergin2004}. The applied dust opacities are taken from \citet{MathisMezgerPanagia1983} and \citet{OssenkopfHenning1994}.

\subsection{Initial conditions and setup parameters}

\begin{figure*}
\begin{minipage}{0.9\textwidth}
\includegraphics[width=\textwidth]{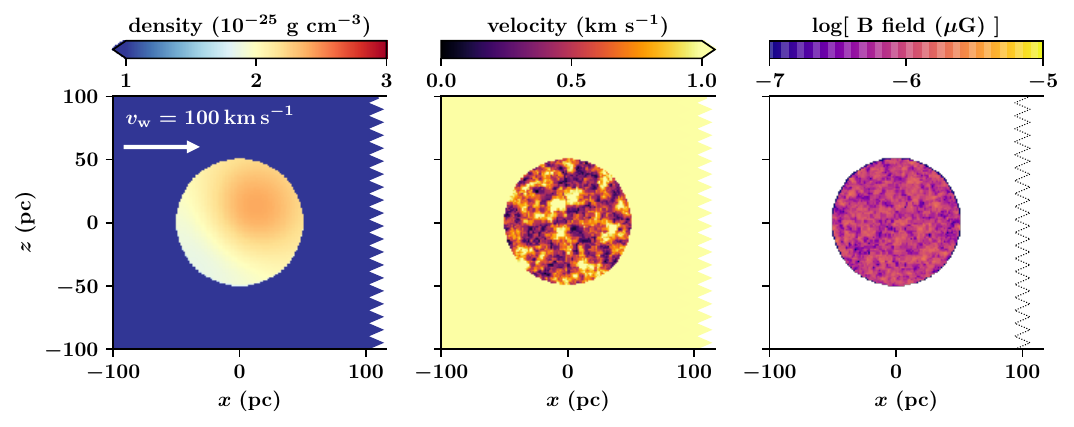}
\caption{Initial conditions for the warm atomic cloud embedded in a hot ionized wind in pressure equilibrium with a small initial density perturbation (left panel). Its turbulent velocity field with a mix of compressive and solenoidal modes has a root mean square speed of $1\,\kmpers$ (middle panle). The tangled magnetic field has a root mean square strength of $1\,\mu\mathrm{G}$ (right panel).}
\label{fig:initial-cloud-configuration}
\end{minipage}
\end{figure*}

\begin{table*}
  \begin{minipage}{0.65\textwidth}
    \centering
    \caption{List of cloud properties. All clouds are exposed to a wind with a velocity of $100\,\kmpers$ and a gas sound speed $c_\mathrm{s,w}=96.6\,\kmpers$.}
    \label{tab:cloud-properties}
    \begin{tabular}{lcccc}
    quantity & units & low-$n$ cloud & medium-$n$ cloud & high-$n$ cloud\\
    \hline	
    number density & $\mathrm{cm^{-3}}$ & 0.1 & 0.5 & 1.0\\
    density              & $\mathrm{g\,cm}^{-3}$ & $2.12\times10^{-25}$ & $1.06\times10^{-24}$ & $2.12\times10^{-24}$\\
    density contrast ($\chi$) & & 100 & 500 & 1000\\
    temperature & K & $7000$ & $1400$ & $700$\\
    plasma $\beta=P_\mathrm{th}/P_\mathrm{mag}$ & & 3.47 & 3.47 & 3.47\\
    cloud sound speed, $c_\mathrm{s}$ & $\kmpers$ & 6.75 & 3.02 & 2.13 \\
    wind sound speed, $c_\mathrm{s,w}$ & $\kmpers$ & 96.6 & 96.6 & 96.6 \\
    sonic Mach number, $\mathcal{M}=v_\mathrm{w}/c_\mathrm{s}$ & & 14.8 & 33.1 & 46.9 \\
    cloud Alfv\'{e}n speed, $v_\mathrm{A}$ & $\kmpers$ & 6.12 & 2.74 & 1.94 \\
    wind Alfv\'{e}n speed, $v_\mathrm{A,w}$ & $\kmpers$ & 73.3 & 73.3 & 73.3 \\
    Alfv\'{e}n Mach number, $\mathcal{M}=v_\mathrm{w}/v_\mathrm{A}$ & & 16.3 & 36.5 & 51.7 \\
    \hline
    masses & & & & \\
    \hline
    cloud mass, $M_\mathrm{c}$ & M$_\odot$ & 1640 & 8200 & 16420\\
    Jeans mass, $M_\mathrm{J}$ & M$_\odot$ & $1.02\times10^7$ & $4.07\times10^5$ & $1.02\times10^5$ \\
    $M_\mathrm{c}/M_\mathrm{J}$ & & $1.61\times10^{-4}$ & 0.02 & 0.16 \\
    KHI mass, $M_\mathrm{KHI}$ & M$_\odot$ & $7.30\times10^{6}$ & $1.46\times10^{6}$ & $0.73\times10^{6}$\\
    $M_\mathrm{c}/M_\mathrm{KHI}$ & & $2.25\times10^{-4}$ & $5.26\times10^{-3}$ & $2.25\times10^{-2}$ \\
    \hline
    time scales & & & &\\
    \hline
    cloud crushing time, $t_\mathrm{cc}$ & Myr & 4.89 & 10.9 & 15.5 \\
    drag time, $t_\mathrm{drag}$ ($B=0$) & Myr & 48.9 & 245 & 489 \\
    drag time, $t_\mathrm{drag,mag}$ ($B=1\,\mu\mathrm{G}$) & Myr & 31.8 & 159 & 318 \\
    free-fall time, $t_\mathrm{ff}$ & Myr & 145 & 64.6 & 45.7 \\
    H$_2$ formation time, $t_\mathrm{H_2}$ & Gyr & 15.0 & 3.0 & 1.5\\
    growth time, $t_\mathrm{grow}$ & Gyr & 0.98 & 4.02 & 7.37
    \end{tabular}
  \end{minipage}
\end{table*}

\begin{table*}
  \begin{minipage}{0.8\textwidth}
    \caption{List of performed simulations}
    \label{tab:simulation-overview}
    \begin{tabular}{l ccccccccc}
      Name   & $B_0$ &  $B$ dir. & eff. res. & $\Delta x$ & $t_\mathrm{sim}$ & $n_\mathrm{c}$ & $n_\mathrm{c}/n_\mathrm{w}$ & self-\\
      & ($\mu$G) & &      & (pc)       & (Myr) & (cm$^{-3}$) & & gravity\\
      \hline
      \multicolumn{9}{l}{Simulations including wind}&\\
      \hline
$\bm{B}=\bm{0},\,n_\mathrm{c}=0.1\,\mathrm{cm}^{-3},\,-\mathrm{sg}$                   & $0$ & -- & $1280\times256\times256$ & $0.78$ & 40 & $0.1$ & $\phantom{1}100$ & no\\
$\bm{B}=\bm{0},\,n_\mathrm{c}=0.1\,\mathrm{cm}^{-3},\,+\mathrm{sg}$                   & $0$ & -- & $1280\times256\times256$ & $0.78$ & 40 & $0.1$ & $\phantom{1}100$ & yes\\
$\bm{B}=\bm{0},\,n_\mathrm{c}=0.5\,\mathrm{cm}^{-3},\,-\mathrm{sg}$                   & $0$ & -- & $1280\times256\times256$ & $0.78$ & 40 & $0.5$ & $\phantom{1}500$ & no\\
$\bm{B}=\bm{0},\,n_\mathrm{c}=0.5\,\mathrm{cm}^{-3},\,+\mathrm{sg}$                   & $0$ & -- & $1280\times256\times256$ & $0.78$ & 40 & $0.5$ & $\phantom{1}500$ & yes\\
$\bm{B}=\bm{0},\,n_\mathrm{c}=1\,\mathrm{cm}^{-3},\,-\mathrm{sg}$                     & $0$ & -- & $1280\times256\times256$ & $0.78$ & 40 & $1.0$ & $1000$ & no\\
$\bm{B}=\bm{0},\,n_\mathrm{c}=1\,\mathrm{cm}^{-3},\,+\mathrm{sg}$                     & $0$ & -- & $1280\times256\times256$ & $0.78$ & 40 & $1.0$ & $1000$ & yes\\
$\bm{B}\parallel\bm{v}_\mathrm{w},\,n_\mathrm{c}=0.1\,\mathrm{cm}^{-3},\,-\mathrm{sg}$ & $1$ & x & $1280\times256\times256$ & $0.78$ & 40 & $0.1$ & $\phantom{1}100$ & no\\
$\bm{B}\parallel\bm{v}_\mathrm{w},\,n_\mathrm{c}=0.1\,\mathrm{cm}^{-3},\,+\mathrm{sg}$ & $1$ & x & $1280\times256\times256$ & $0.78$ & 40 & $0.1$ & $\phantom{1}100$ & yes\\
$\bm{B}\parallel\bm{v}_\mathrm{w},\,n_\mathrm{c}=0.5\,\mathrm{cm}^{-3},\,-\mathrm{sg}$ & $1$ & x & $1280\times256\times256$ & $0.78$ & 40 & $0.5$ & $\phantom{1}500$ & no\\
$\bm{B}\parallel\bm{v}_\mathrm{w},\,n_\mathrm{c}=0.5\,\mathrm{cm}^{-3},\,+\mathrm{sg}$ & $1$ & x & $1280\times256\times256$ & $0.78$ & 40 & $0.5$ & $\phantom{1}500$ & yes\\
$\bm{B}\parallel\bm{v}_\mathrm{w},\,n_\mathrm{c}=1\,\mathrm{cm}^{-3},\,-\mathrm{sg}$   & $1$ & x & $1280\times256\times256$ & $0.78$ & 40 & $1.0$ & $1000$ & no\\
$\bm{B}\parallel\bm{v}_\mathrm{w},\,n_\mathrm{c}=1\,\mathrm{cm}^{-3},\,+\mathrm{sg}$   & $1$ & x & $1280\times256\times256$ & $0.78$ & 40 & $1.0$ & $1000$ & yes\\
$\bm{B}\perp\bm{v}_\mathrm{w},\,n_\mathrm{c}=0.1\,\mathrm{cm}^{-3},\,-\mathrm{sg}$     & $1$ & y & $1280\times256\times256$ & $0.78$ & 40 & $0.1$ & $\phantom{1}100$ & no\\
$\bm{B}\perp\bm{v}_\mathrm{w},\,n_\mathrm{c}=0.1\,\mathrm{cm}^{-3},\,+\mathrm{sg}$     & $1$ & y & $1280\times256\times256$ & $0.78$ & 40 & $0.1$ & $\phantom{1}100$ & yes\\
$\bm{B}\perp\bm{v}_\mathrm{w},\,n_\mathrm{c}=0.5\,\mathrm{cm}^{-3},\,-\mathrm{sg}$     & $1$ & y & $1280\times256\times256$ & $0.78$ & 40 & $0.5$ & $\phantom{1}500$ & no\\
$\bm{B}\perp\bm{v}_\mathrm{w},\,n_\mathrm{c}=0.5\,\mathrm{cm}^{-3},\,+\mathrm{sg}$     & $1$ & y & $1280\times256\times256$ & $0.78$ & 30 & $0.5$ & $\phantom{1}500$ & yes\\
$\bm{B}\perp\bm{v}_\mathrm{w},\,n_\mathrm{c}=1\,\mathrm{cm}^{-3},\,-\mathrm{sg}$       & $1$ & y & $1280\times256\times256$ & $0.78$ & 20 & $1.0$ & $1000$ & no\\
$\bm{B}\perp\bm{v}_\mathrm{w},\,n_\mathrm{c}=1\,\mathrm{cm}^{-3},\,+\mathrm{sg}$ & $1$ & y & $1280\times256\times256$ & $0.78$ & 40 & $1.0$ & $1000$ & yes\\
      \hline
      \multicolumn{9}{l}{Simulations without wind}&\\
      \hline
      noW, $n_\mathrm{c}=0.1\,\mathrm{cm}^{-3},\,-\mathrm{sg}$ & -- & -- & $256\times256\times256$ & $0.78$ & 40 & $0.1$ & $\phantom{1}100$ & no\\
      noW, $n_\mathrm{c}=0.1\,\mathrm{cm}^{-3},\,+\mathrm{sg}$ & -- & -- & $256\times256\times256$ & $0.78$ & 40 & $0.1$ & $\phantom{1}100$ & yes\\
      noW, $n_\mathrm{c}=0.5\,\mathrm{cm}^{-3},\,-\mathrm{sg}$ & -- & -- & $256\times256\times256$ & $0.78$ & 40 & $0.5$ & $\phantom{1}500$ & no\\
      noW, $n_\mathrm{c}=0.5\,\mathrm{cm}^{-3},\,+\mathrm{sg}$ & -- & -- & $256\times256\times256$ & $0.78$ & 40 & $0.5$ & $\phantom{1}500$ & yes\\
      noW, $n_\mathrm{c}=1\,\mathrm{cm}^{-3},\,-\mathrm{sg}$ & -- & -- & $256\times256\times256$ & $0.78$ & 40 & $1.0$ &                     $1000$ & no\\
      noW, $n_\mathrm{c}=1\,\mathrm{cm}^{-3},\,+\mathrm{sg}$ & -- & -- & $256\times256\times256$ & $0.78$ & 40 & $1.0$ &                    $1000$ & yes\\

    \end{tabular}

    \medskip
	Listed are the simulation name, magnetic field strength in the wind ($B_0$), the field orientation in the wind ($B$ dir.), the effective resolution, the corresponding maximum spatial resolution ($\Delta x$), the simulated time ($t_\mathrm{sim}$), the cloud number density ($n_\mathrm{c}$), the density contrast between the cloud and the ambient wind ($n_\mathrm{c}/n_\mathrm{w}$), and whether self-gravity is included or not. 
  \end{minipage}
\end{table*}

We set up an elongated box with a size of $1000\times200\times200\,\mathrm{pc}^3$, in which we place a spherical cloud of warm gas at the left edge of the box as illustrated in Fig.~\ref{fig:initial-cloud-configuration}. The hot ambient gas surrounding the cloud has a uniform total number density of $n_\mathrm{w}=10^{-3}\,\mathrm{cm}^{-3}$ for all simulations and a temperature of $10^6\,\mathrm{K}$, and is fully ionized. The total mass of the background gas integrates to $870\,\mathrm{M}_\odot$. The cloud with a radius of $50\,\mathrm{pc}$ is in pressure equilibrium with the ambient medium with all hydrogen in the neutral atomic phase. For the mean number density of the cloud we choose $n_\mathrm{c}=0.1$, $0.5$ and $1\,\mathrm{cm}^{-3}$, which we refer to as \emph{low}, \emph{medium}, and \emph{high} density cloud. The three configurations correspond to a density contrast $\chi=\rho_\mathrm{c}/\rho_\mathrm{w}\approx n_\mathrm{c}/n_\mathrm{w}$ of $100$, $500$, and $1000$. The resulting cloud masses are $1640$, $8200$, and $16400\,\mathrm{M}_\odot$; the cloud temperatures read $7000$, $1400$, and $700\,\mathrm{K}$, respectively. The density of the cloud is close to uniform with a small asymmetric density perturbation with respect to the cloud as indicated in the left-hand panel of Fig~\ref{fig:initial-cloud-configuration}. This small over-density helps breaking the symmetry of the cloud in the further evolution. We stress that the detailed shape of the asymmetry does not matter in the presence of the turbulent velocities that we add.

As an indicator for how gravitationally unstable the initial clouds are, we compute the Jeans mass of the gas in the cloud, $M_\mathrm{J}$, as a sphere with a diameter of the Jeans length \citep{Jeans1902},
\begin{equation}
M_\mathrm{J} = \frac{4\pi}{3}\rho_\mathrm{c}\left(\frac{\lambda_\mathrm{J}}{2}\right)^3,
\end{equation}
with the density $\rho_\mathrm{c}$ and the Jeans length, $\lambda_\mathrm{J}$,
\begin{equation}
\lambda_\mathrm{J} = \sqrt{\frac{c_\mathrm{s}^2}{G\rho_c}}.
\end{equation}
Here the sound speed is $c_\mathrm{s}=\sqrt{\gamma k_\mathrm{B} T_\mathrm{c}/(\mu m_\mathrm{p})}$ with the adiabatic index $\gamma=5/3$, the Boltzmann constant $k_\mathrm{B}$, the cloud temperature $T_\mathrm{c}$, the gravitational constant $G$, the mean molecular weight $\mu$, and the proton mass $m_\mathrm{p}$. The Jeans masses for the three cloud setups are $M_\mathrm{J}(n_\mathrm{c}=0.1\,\mathrm{cm}^{-3}) \approx 10^7\,\mathrm{M}_\odot$, $M_\mathrm{J}(n_\mathrm{c}=0.5\,\mathrm{cm}^{-3}) \approx 4\times10^5\,\mathrm{M}_\odot$, and $M_\mathrm{J}(n_\mathrm{c}=1\,\mathrm{cm}^{-3}) \approx 10^5\,\mathrm{M}_\odot$, respectively. This means that in all cases the clouds are initially supported by thermal pressure against gravitational collapse.

We also compare the cloud masses to the mass above which clouds should be stabilized against the Kelvin-Helmholtz instability by its self-gravity \citep{MurrayEtAl1993},
\begin{equation}
M_\mathrm{KHI} = \frac{2\upi r_\mathrm{c} v_\mathrm{w}^2}{G\chi},
\end{equation}
which is significantly larger than the cloud masses. The ratios $M_\mathrm{c}/M_\mathrm{KHI}$ read $2.2\times10^{-4}$, $5.26\times10^{-3}$, and $2.25\times10^{-2}$, so all clouds are expected to be dissolved by the Kelvin-Helmholtz instability. An overview of the most important parameters is given in table~\ref{tab:cloud-properties}. The time scales are discussed separately in Section~\ref{sec:char-numbers}.

In addition to the density perturbation we apply a small turbulent velocity field in the cloud. The motions are created in Fourier space with a three-dimensional turbulent power spectrum $P(k)\propto k^{-4}$, where $k=m\pi/r_\mathrm{c}$ is the modulus of the wave vector and the exponent of $-4$ corresponds to Burgers turbulence \citep{Burgers1948}. The minimum $k$ vector has $m_\mathrm{min}=4$, i.e. the largest spatial mode spans one quarter of the cloud. The smallest spatial scales of the turbulence are at the resolution limit. The root mean square velocity of the turbulent motions is $1\,\mathrm{km\,s}^{-1}$. The cloud is also permeated by a magnetic field with a root mean square field strength of a $1\,\mu\mathrm{G}$. The tangled field is generated in a similar fashion as the turbulent velocity profile with a power spectrum of $P(k)\propto k^{-4}$. The divergence-free condition is achieved by projection of the Fourier modes into solely rotational modes. For the minimum mode we apply again $m_\mathrm{min}=4$. The initial density, velocity and magnetic field configuration is shown in Fig.~\ref{fig:initial-cloud-configuration}.

The cloud is exposed to a hot wind entering the simulation box at the left-hand $x$ boundary as indicated in Fig.~\ref{fig:initial-cloud-configuration}. The injected wind temperature and density are identical to the ambient medium. The wind velocity is set to $100\,\kmpers$ for all runs. For the right-hand edge in $x$ as well as both sides of the $y$ and $z$ direction we apply diode boundary condition. These are boundaries which allow the gas to leave the box but not enter. We distinguish between three different wind setups, namely a non-magnetic wind as well as two magnetised counterparts with the magnetic field of the wind being orientated along $x$ and along $y$, i.e. one configuration in which the field is parallel to the wind direction and one with the field perpendicular to the flow. In the following we refer to the three configurations as \emph{non-magnetic}, \emph{parallel field}, and \emph{perpendicular field}, respectively. The motivation for the three wind configurations is the following. The field strength in the hot regions of the ISM and in the circum-galactic medium can vary noticeably \citep[e.g.][]{Beck2015, Han2017}. Simulations suggest that the field in hot expanding bubbles can be very weak due to strong expansion and the resulting adiabatic weakening \citep[e.g.][]{GirichidisEtAl2018a}, which motivates the non-magnetic setup. The correlation length of the field in spiral galaxies is typically larger than the modelled cloud with a diameter of 100\,pc. The field in spiral galaxies is often parallel to the disc, however, there is a variety of local variations in the field configuration \citep[e.g.][]{Haverkorn2015}. Consequently, if a cloud is hit by a large-scale coherent outflow from the disc, the field is likely to be perpendicular to the outflow. Contrary, if the cloud forms or moves along opened up field lines -- see e.g. the X-shaped field structures -- the hot wind is more likely to move parallel to the field lines.
We further differentiate between runs with and without the impact of self-gravity. Finally, we also run six comparison simulations without a wind; three different cloud densities with and without self-gravity each. The clouds in isolation enable us to directly compare how much compression, cooling and molecular gas is triggered, accelerated or prevented by the wind. An overview of all simulations and chosen combinations is given in table~\ref{tab:simulation-overview}. We simulate most of the runs for $40\,\mathrm{Myr}$ except for runs $\bm{B}\perp\bm{v}_\mathrm{w},\,n_\mathrm{c}=0.5\,\mathrm{cm}^{-3},\,+\mathrm{sg}$ and $\bm{B}\perp\bm{v}_\mathrm{w},\,n_\mathrm{c}=1\,\mathrm{cm}^{-3},\,-\mathrm{sg}$ which are terminated at $30$ and $20\,\mathrm{Myr}$, respectively, because of prohibitively small time steps. In both cases the perpendicular field configuration causes strong fields in regions of very low density, which leads to high Alfv\'{e}n speeds and thus small time steps.

\section{Characteristic numbers}
\label{sec:char-numbers}

The growth, survival, and destruction of clouds can be illustrated by a comparison between different time scales. The destruction of cold clouds in a hot wind is given by the \emph{cloud crushing time} \citep{KleinMcKeeColella1994},
\begin{equation}
t_\mathrm{cc} = \chi^{1/2} \frac{r_\mathrm{c}}{v_\mathrm{w}} = \sqrt{\frac{\rho_\mathrm{c}}{\rho_\mathrm{w}}}\frac{r_\mathrm{c}}{v_\mathrm{w}},
\end{equation}
where $r_\mathrm{c}$ and $\rho_\mathrm{c}$ are the cloud radius and mass density. The wind velocity and density are $v_\mathrm{w}$ and $\rho_\mathrm{w}$. The density contrast is $\chi=\rho_\mathrm{c}/\rho_\mathrm{w}$. For our choice of parameters the cloud crushing time scale ranges between $5$ and $16\,\mathrm{Myr}$. Simulations of clouds using an adiabatic equation of state find that clouds are destroyed by hydrodynamic instabilities, in particular the Kelvin-Helmholtz and Rayleigh-Taylor instabilities, within a few $t_\mathrm{cc}$ \citep{KleinMcKeeColella1994, XuStone1995, NakamuraEtAl2006, BandaBarraganEtAl2016, PittardParkin2016, GoldsmithPittard2017, GoldsmithPittard2018}. We note that during the initial interaction of the wind with the cloud and the resulting shock, the Richtmyer-Meshkov instability \citep{Richtmyer1960, Meshkov1969, Zhou2017a} can be triggered. However, the clouds evolution mainly depends on the continuous impact of the wind rather than the initial shock.

A measure for the acceleration of clouds is given by the \emph{drag time scale}
\begin{equation}
t_\mathrm{drag} = \chi \frac{r_\mathrm{c}}{v_\mathrm{w}} = \chi^{1/2}\, t_\mathrm{cc},
\end{equation}
indicating that the clouds are likely to be destroyed before they can be accelerated for density contrasts $\chi\gg 1$. The non-magnetic values ($t_\mathrm{drag}$) for our setups span a range from $\sim50$ to $\sim500\,\mathrm{Myr}$. The efficiency of the acceleration is enhanced by magnetic fields with a modified drag force and a resulting time scale \citep{McCourtEtAl2015}
\begin{equation}
t_\mathrm{drag,mag} = \chi^{1/2}\, t_\mathrm{cc}\,\frac{v_\mathrm{w}^2}{v_\mathrm{w}^2+v_\mathrm{a}^2},
\end{equation}
with the Alfv\'{e}n speed $v_\mathrm{a}=B/\sqrt{4\pi\rho}$. In our case this reduces the drag times to values between $31.8$ and $318\,\mathrm{Myr}$, i.e. a reduction by 35 per cent compared to the non-magnetic case.

Radiative cooling provides a powerful mechanism to allow for a contraction of the cloud and thus a reduction of the effective cross section, so the destruction of clouds would be slowed down by suppressing the Kelvin-Helmholtz and Rayleigh-Taylor instabilities. Simulations of clouds including cooling support this paradigm \citep{MellemaKurkRoettgering2002,CooperEtAl2009, ScannapiecoBrueggen2015, GronnowEtAl2018, SparrePfrommerVogelsberger2019}, but highlight that the clouds are effectively destroyed anyway. While cooling prolongs the life time of the clouds, the resulting smaller cross section impedes the effective acceleration \citep{ScannapiecoBrueggen2015, SchneiderRobertson2017}. The gas cooling time can be approximated by
\begin{equation}
t_\mathrm{cool,c} \sim \frac{u}{\dot{u}} = \frac{nk_\mathrm{B}T}{n_\mathrm{c}^2\Lambda}
\end{equation}
with the cooling function $\Lambda$. For temperatures in the range of $10^3$ to $10^4\,\mathrm{K}$ and cooling rates between $\Lambda\sim10^{-26}-10^{-24}\,\mathrm{erg\,s^{-1}\,cm^3}$ \citep{DalgarnoMcCray1972} we find cooling times between $0.05$ and $5\,\mathrm{Myr}$, see also \citep{KlessenGlover2016, GirichidisEtAl2020b}.

\citet{GronkeOh2018} and \citet{GronkeOh2020a} investigate the growth of clouds due to the wind via mixing of wind and cloud material combined with radiative cooling. They find that gas mixing is mainly driven by radiative cooling rather than hydrodynamic instabilities in agreement with \citet{JiOhMasterson2019} and derive the \emph{growth time}, 
\begin{equation}
t_\mathrm{grow} \sim\chi t_\mathrm{sc,c}\left(\frac{t_\mathrm{cool,c}}{t_\mathrm{sc,c}}\right)^{1/4}
\end{equation}
with the cooling time of the cloud $t_\mathrm{cool,c}$ and the sound crossing time of the cloud $t_\mathrm{sc,c}$. For our setups the sound crossing times range from $15$ to $45\,\mathrm{Myr}$ and the resulting growth times are between $\sim1$ and $8\,\mathrm{Gyr}$.

Most of the previous studies focused on clouds that are still dilute enough, so that self-gravity does not play a role. However, in the case of molecular gas the densities are typically large enough for self-gravity to be important. We thus also mention the \emph{free-fall time}
\begin{equation}
t_\mathrm{ff} = \sqrt{\frac{3\pi}{32G\rho_\mathrm{c}}}.
\end{equation}
The initial densities of the clouds yield values from $46$ to $145\,\mathrm{Myr}$. 

Finally, we quote an estimate for the formation time scale of molecular hydrogen \citep{HollenbachWernerSalpeter1971},
\begin{equation}
t_\mathrm{H_2} = \frac{1.5\,\mathrm{Gyr}}{n_\mathrm{c}/1\,\mathrm{cm}^{-3}}.
\end{equation}
Given the initial densities of the clouds, $t_\mathrm{H_2}$ is $1.5$ to $15\,\mathrm{Gyr}$, which is orders of magnitude longer than the simulated time scales. The formation of molecular gas thus relies on efficient compression of the clouds by the wind and/or self-gravity.

We simulate the systems for $40\,\mathrm{Myr}$. Since the shortest time scale is the cloud crushing time with only a fraction of the simulated time we naively expect the cloud to be destroyed. During this time we would not expect to form a noticeable amount of molecular gas without the wind or self-gravity.

\section{Morphological evolution}
\label{sec:morphology}

\begin{figure*}
\begin{minipage}{0.9\textwidth}
\includegraphics[width=\textwidth]{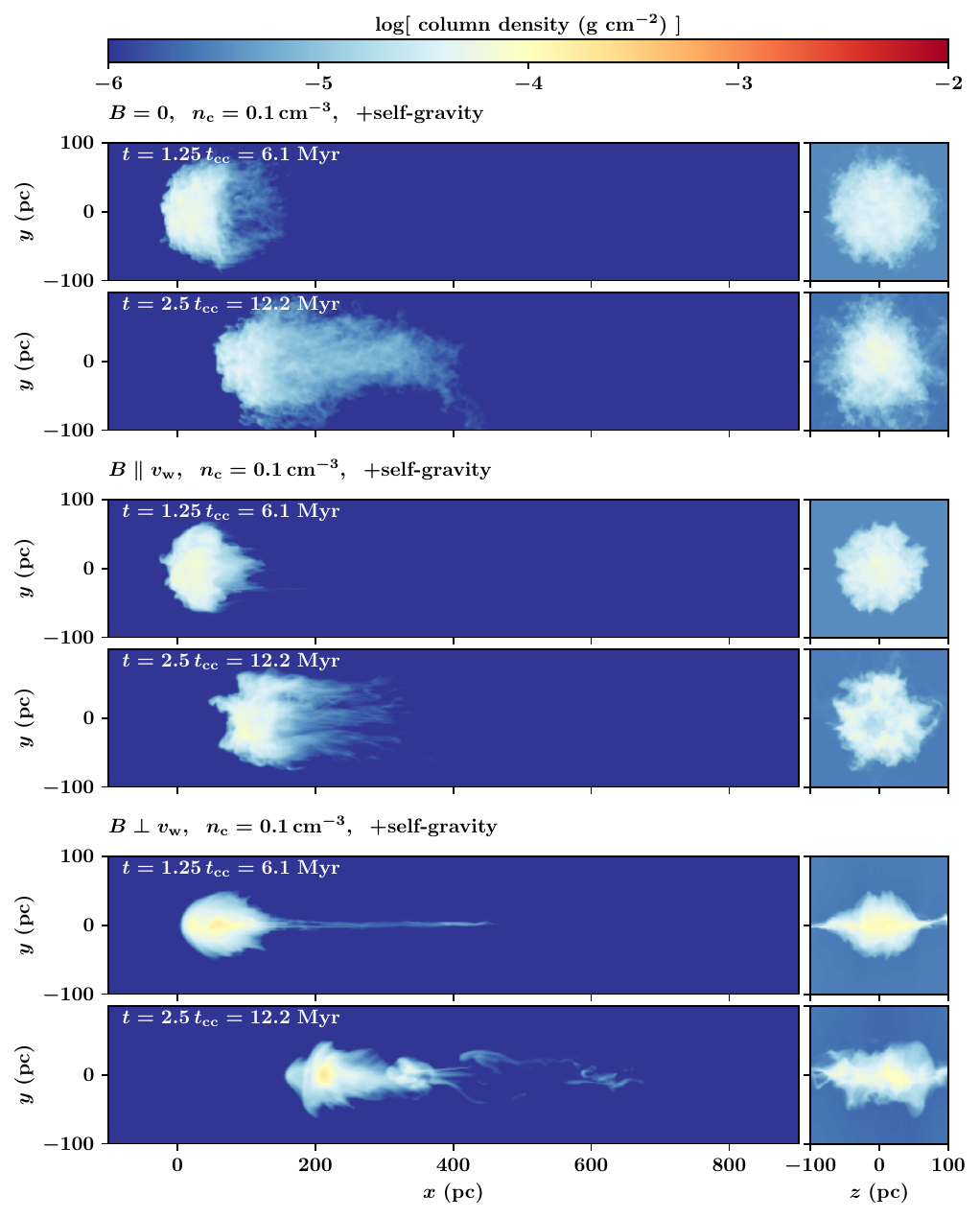}
\caption{Column density of all simulations with an initial cloud density of $n_\mathrm{c}=0.1\,\mathrm{cm^{-3}}$ and self-gravity at $t=1.25\,t_\mathrm{cc}=6.1\,\mathrm{Myr}$ and $t=2.5\,t_\mathrm{cc}=12.2\,\mathrm{Myr}$. From top to bottom we show the non-magnetic wind, the magnetized wind with $B$ along the flow, and the wind with $B$ perpendicular to the flow. The wind with the perpendicular field can keep the cloud confined, compresses it to reach the highest column density and can accelerate it most efficiently.}
\label{fig:cd-time-L6-d010-sg}
\end{minipage}
\end{figure*}

\begin{figure*}
\begin{minipage}{0.9\textwidth}
\includegraphics[width=\textwidth]{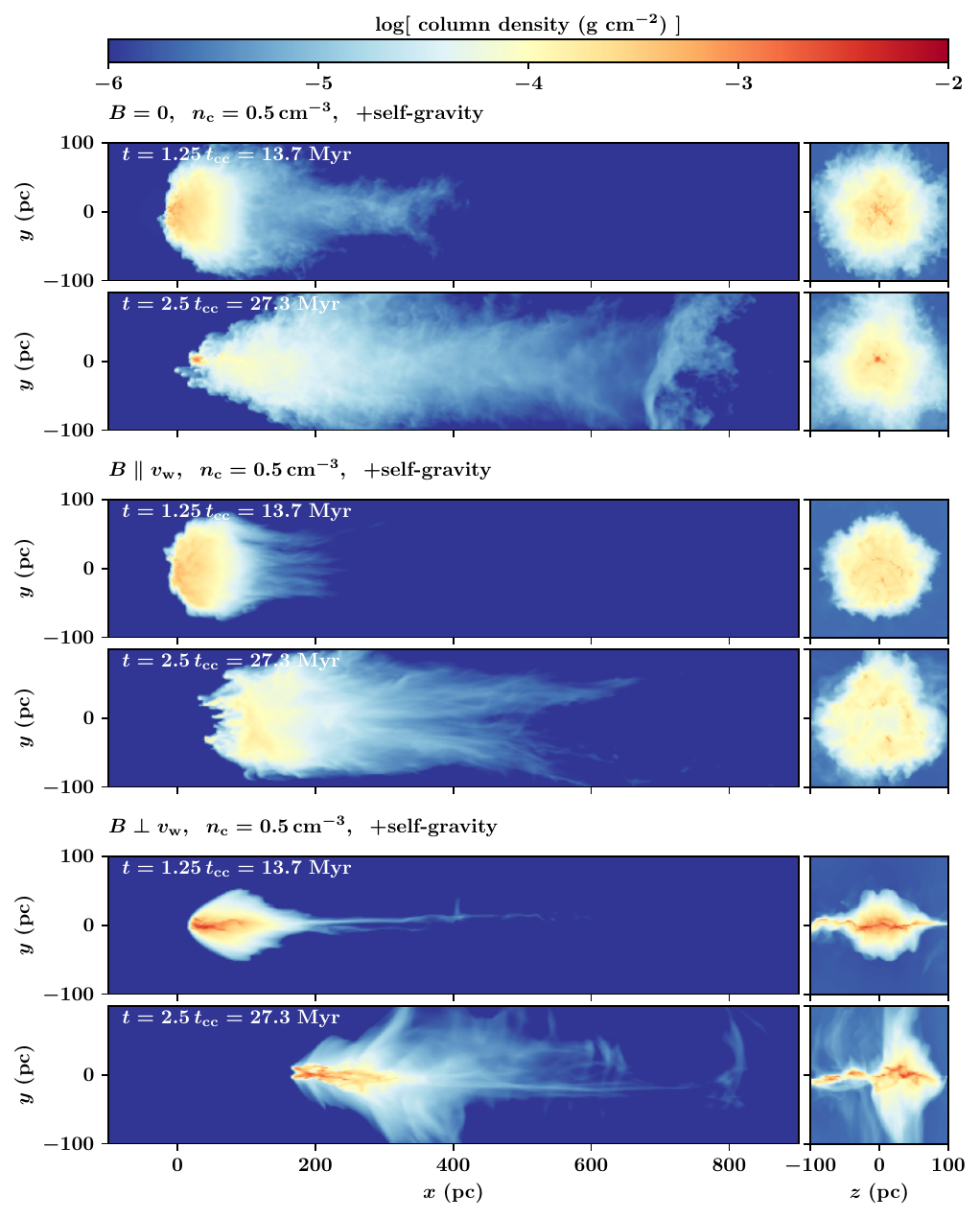}
\caption{Column density of all simulations with an initial cloud density of $n_\mathrm{c}=0.5\,\mathrm{cm^{-3}}$ and self-gravity at $t=1.25\,t_\mathrm{cc}=13.7\,\mathrm{Myr}$ and $t=2.5\,t_\mathrm{cc}=27.3\,\mathrm{Myr}$. From top to bottom we show the non-magnetic wind, the magnetized wind with $B$ along the flow, and the wind with $B$ perpendicular to the flow. Only the simulation with $\Bperpv$ forms a large coherent region of dense gas. The other two setups only form small high-density fragments.}
\label{fig:cd-time-L6-d050-sg}
\end{minipage}
\end{figure*}

\begin{figure*}
\begin{minipage}{0.9\textwidth}
\includegraphics[width=\textwidth]{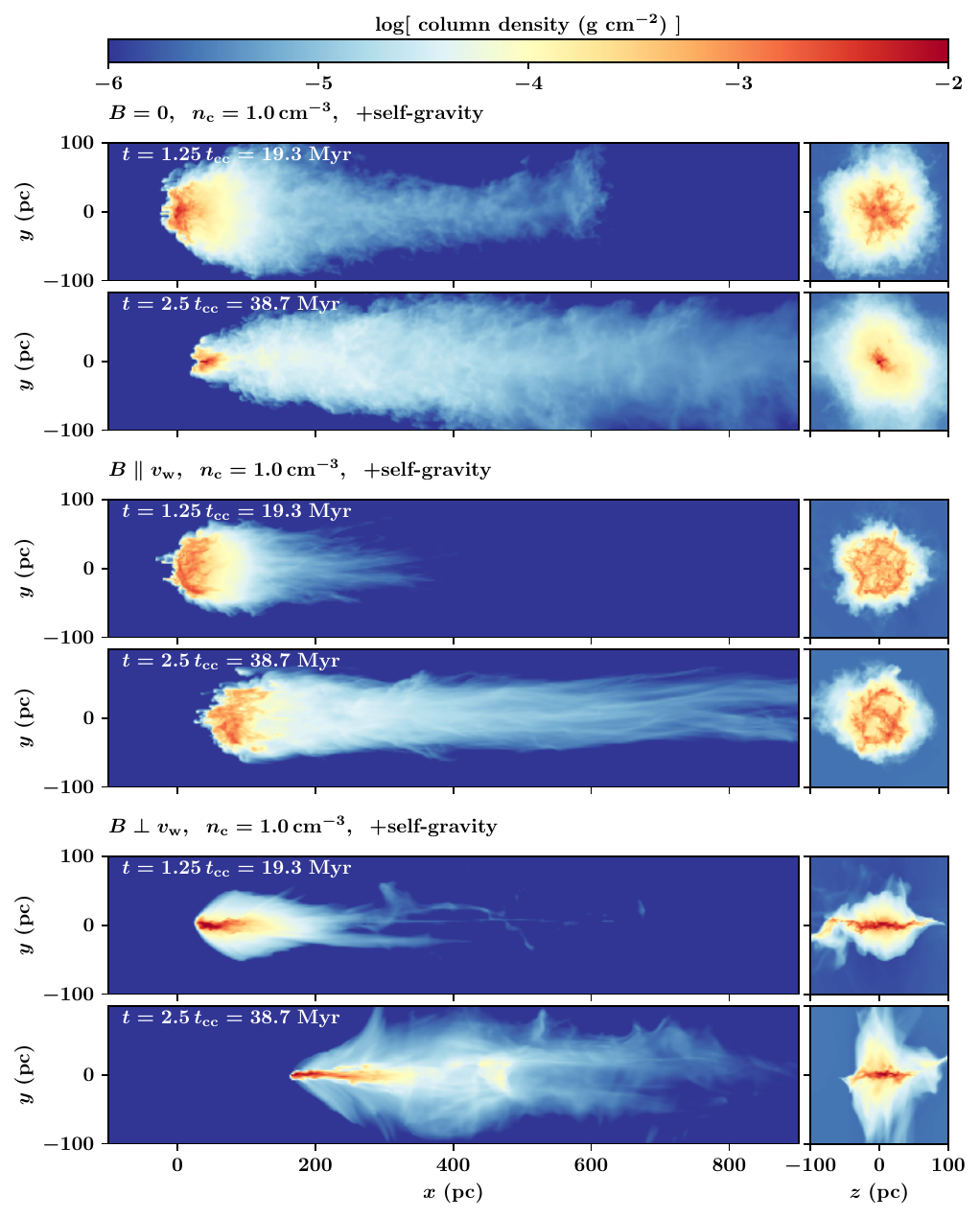}
\caption{Column density of all simulations with an initial cloud density of $n_\mathrm{c}=1\,\mathrm{cm^{-3}}$ and self-gravity at $t=1.25\,t_\mathrm{cc}=19.3\,\mathrm{Myr}$ and $t=2.5\,t_\mathrm{cc}=38.7\,\mathrm{Myr}$. From top to bottom we show the non-magnetic wind, the magnetized wind with $B$ along the flow, and the wind with $B$ perpendicular to the flow. For $\Bparv$ the compression perpendicular to the field is suppressed by the magnetic pressure. As a result the dense gas forms in numerous small fragments instead of one massive central over-density as in the non-magnetic or the $\Bperpv$ setup.}
\label{fig:cd-time-L6-d100-sg}
\end{minipage}
\end{figure*}

\begin{figure*}
\begin{minipage}{\textwidth}
\includegraphics[width=\textwidth]{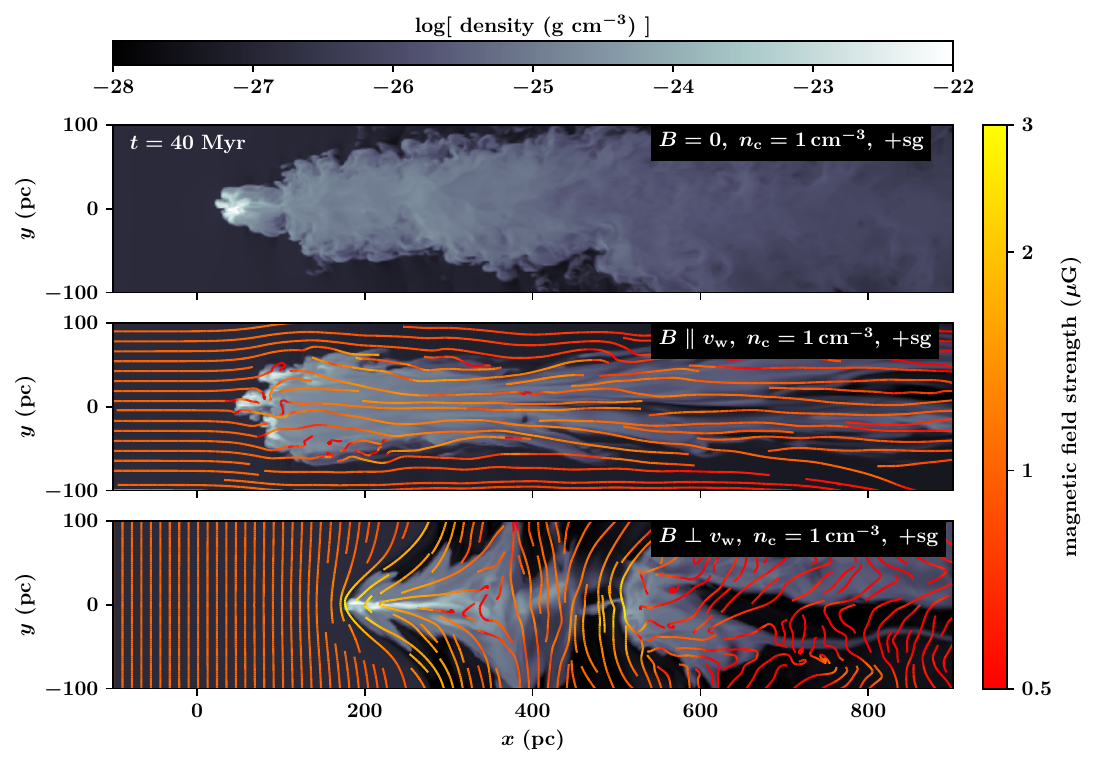}
\caption{Density cuts and magnetic field stream lines for the simulations with the densest gas clouds including self-gravity at the end of the simulation time at $t=40\,\mathrm{Myr}$. The streamlines are coloured with the magnetic field strength. The magnetic wind with a field parallel to the wind velocity prevents the clouds from contracting perpendicular to the field lines, which is a natural consequence of the ideal MHD approximation. In the case of a perpendicular field the draping of the field lines does not only compress the cloud along the $y$ direction but also more efficiently accelerates the gas with the wind. The field strength at the tip of the cloud is enhanced in this case by a factor of a few.}
\label{fig:dens-slice-mag-strm}
\end{minipage}
\end{figure*}

The morphological evolution differs early on between the individual setups. An overview of the column densities for all simulations including self-gravity is depicted in Fig.~\ref{fig:cd-time-L6-d010-sg}, \ref{fig:cd-time-L6-d050-sg} and~\ref{fig:cd-time-L6-d100-sg} for the three different initial density contrasts of the clouds (the comparison simulations without the wind are presented in Appendix~\ref{sec:nowind}). Each figure consists of three different groups, which correspond to the different types of the wind and illustrate the structure of the cloud at two different simulation times. The left-hand elongated panels show the clouds from the side, the right-hand panels are the view along with the wind direction and illustrate the effective cross section of the gas cloud with the wind. We show the column densities after $1.25$ and $2.5\,t_\mathrm{cc}$. In Fig.~\ref{fig:cd-time-L6-d010-sg} (low density cloud, $n_\mathrm{c}=0.1\,\percc$) this corresponds to $6.1$ and $12.2\,\mathrm{Myr}$. The clouds with medium density ($n_\mathrm{c}=0.5\,\percc$) are shown at $13.7$ and $27.3\,\mathrm{Myr}$ (Fig.~\ref{fig:cd-time-L6-d050-sg}), and the clouds with the high density ($n_\mathrm{c}=1\,\percc$) are shown at $19.3$ and $38.7\,\mathrm{Myr}$ in Fig.~\ref{fig:cd-time-L6-d100-sg}.

Of the parameters that we vary, the main parameter determining the cloud evolution is its initial density. All clouds with initially the low density ($n_\mathrm{c}=0.1\,\percc$, Fig.~\ref{fig:cd-time-L6-d010-sg}) disperse over the course of the simulation. Whether self-gravity is included or not does not alter the results noticeably. The magnetic field impacts the morphology of the cloud but does not prevent the cloud from being disrupted at the end of the simulation. The compression due to the wind is not enough to generate a dense region that can cool sufficiently fast to form clumpy structures that are kept together by gravitational attraction. Nor does a shielded volume form, in which atomic gas can condense to the molecular phase. The maximum visual extinction, $A_\mathrm{v}$ \citep{BohlinSavageDrake1978}, does not exceed 0.07.  

For the medium density clouds ($n_\mathrm{c}=0.5\,\mathrm{cm}^{-3}$, Fig.~\ref{fig:cd-time-L6-d050-sg}) the two other parameters (magnetic field orientation and self-gravity) have a significant impact. In a non-magnetised wind the cloud can form a central dense condensation followed by an extended low-density tail of stripped material. In the magnetised wind with parallel field, the induced perturbations can disperse the cloud and prevent the formation of a central dense region. The cloud quickly breaks into small substructures that are efficiently pushed away by the wind. A magnetic field perpendicular to the flow bends around the overdensity and aids in keeping a coherent gas structure, which can form an elongated cloud along the $z$ direction. Clouds with medium density ($n_\mathrm{c}=0.5\,\mathrm{cm}^{-3}$) form molecular gas within the first $10-15\,\mathrm{Myr}$ in all cases but only for the wind with perpendicular magnetic fields does the fraction of H$_2$ exceed the per cent level with a maximum $A_\mathrm{v}$ of 0.36.

\begin{figure*}
\begin{minipage}{\textwidth}
\includegraphics[width=0.33\textwidth]{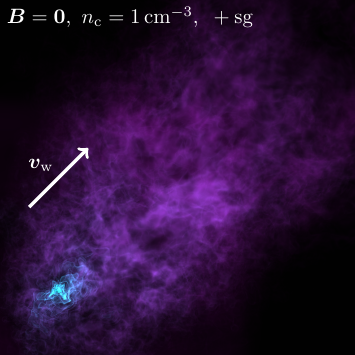}
\includegraphics[width=0.33\textwidth]{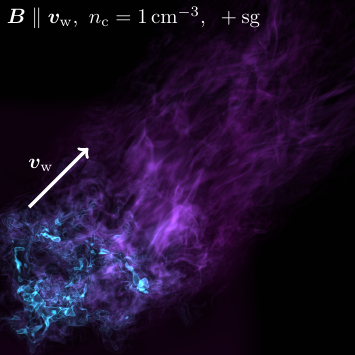}
\includegraphics[width=0.33\textwidth]{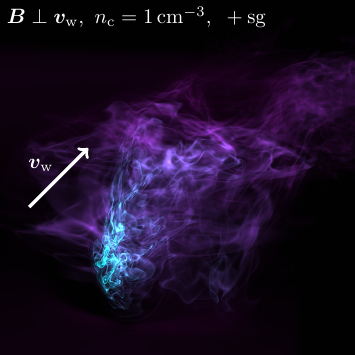}
\caption{Ray-tracing images of the gas density for all runs with $n_\mathrm{c}=1.0\,\mathrm{cm}^{-3}$ at $t=2.5\,t_\mathrm{cc}$. The colors from dark purple to light cyan depict the gas density in the range from $10^{-27}$ to $10^{-22}\,\gpercc$. From left to right we show non-magnetic wind, the wind with $\Bparv$, and $\Bperpv$. Besides the differences in the morphology of the dense cloud, the variations in the tail of the stripped material are illustrated.}
\label{fig:density-ray-tracing}
\end{minipage}
\end{figure*}

In the setups with the high density ($n_\mathrm{c}=1.0\,\mathrm{cm}^{-3}$, Fig.~\ref{fig:cd-time-L6-d100-sg}) all clouds form dense regions and molecular gas inside them, $A_\mathrm{v}>1$. A noticeable difference in the morphology of the condensations is again caused by the magnetic field details. The non-magnetic run forms one concentrated cloud, the setup with $\Bparv$ shows multiple disconnected fragments, and for $\Bperpv$ the field generates an elongated filamentary cloud. The flattened structure is caused by preventing the contraction along $z$ due to the magnetic pressure.  For this cloud density we depict the differences between the wind properties in Fig.~\ref{fig:dens-slice-mag-strm}, where we plot the density in a cut through the centre of the box together with the magnetic field streamlines color coded with the field strength. The non-magnetic wind allows for a compression of the cloud in the central region of the box. This is partially prevented in the case of a parallel field ($\Bparv$) due to the additional magnetic pressure acting perpendicular to the field and the magnetic tension that acts against bending the field lines. The draping of magnetic field lines around the cloud in the case of a perpendicular field ($\Bperpv$) helps to confine the gas near the $y=0$ plane. One can clearly see the differences in the magnetic field strengths between the parallel and perpendicular configurations. Whereas in the former one the field shows weak and random variations in the field strength, the perpendicular configuration shows a significant enhancement at the cloud-wind interface at $x\sim200\,\mathrm{pc}$ as well as a second enhancement at $x\sim500\,\mathrm{pc}$.

We illustrate the morphology of the dense clouds ($n_\mathrm{c}=1\,\percc$) at $t=2.5\,t_\mathrm{cc}$ with ray-tracing images of the gas density in Fig.~\ref{fig:density-ray-tracing}. The wind direction points diagonally from bottom left to top right. From left to right we show the three wind configurations. The light cyan indicates high densities ($n\sim100\,\percc$), the dark purple shows the low density tail ($n\sim10^{-3}\,\percc$). Besides the diversity in the clouds, the images illustrate the different morphology of the cloud and its stripped tail.

\section{Effective acceleration}
\label{sec:acceleration}

\begin{figure}
\centering
\includegraphics[width=8cm]{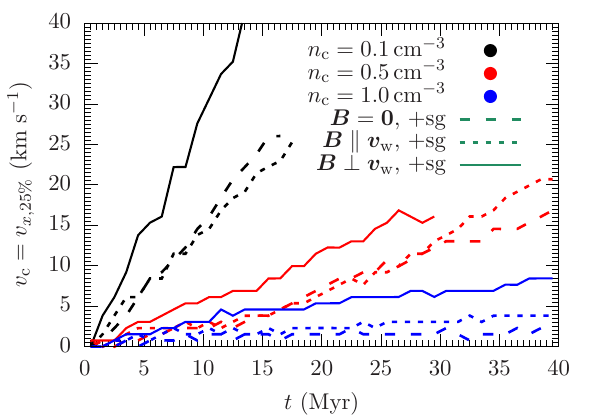}
\caption{Velocity of the cloud in $x$ direction, whose position we set to the 25 percentile mark of the integrated mass along $x$. In all cases the velocity can be well described by a linear function, which reveals a constant acceleration. For all densities the wind with $\Bperpv$ can accelerate the cloud most efficiently.}
\label{fig:velocity-25percent-mark}
\end{figure}

\begin{figure}
\centering
\includegraphics[width=8cm]{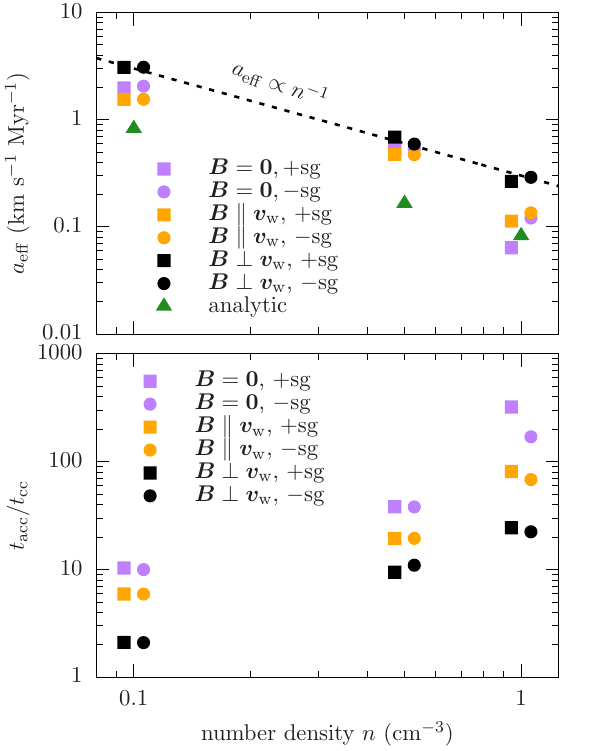}
\caption{Top: Derived accelerations from the velocity of the clouds based on linear fits to the time evolution of the cloud speed (see Fig.~\ref{fig:velocity-25percent-mark}). The expected scaling based on the drag force ($a_\mathrm{eff}\propto n^{-1}$) is indicated by the dashed line and followed closely in the $\Bperpv$ case. Analytic estimates are shown in green. Bottom: Acceleration time scale $t_\mathrm{acc} = v_\mathrm{w}/a_\mathrm{eff}$ in units of the cloud crushing time. For the dense clouds it takes around $20-200\,t_\mathrm{cc}$ to accelerate the cloud. We apply an artificial offset in the abscissa between the simulations with and without self-gravity for better visibility.}
\label{fig:effective-accelerations}
\end{figure}

How strongly the gas is accelerated depends on the gas density. Low-density gas is easily blown away with the wind whereas dense gas only slowly gains speed. We can estimate the effective acceleration using the drag force \citep[e.g.][]{ShuAstroGas1992, Lautrup2011}
\begin{align}
F_\mathrm{d} &= M_\mathrm{c}a_\mathrm{c} =\frac{1}{2}\rho_\mathrm{w}v_\mathrm{w}^2C_\mathrm{d}A,
\end{align}
where $M_\mathrm{c}$ and $a_\mathrm{c}$ are the mass and the acceleration of the cloud, $\rho_\mathrm{w}$ is the density of the wind, $C_\mathrm{d}$ is the drag coefficient and $A$ is the effective cross section. The drag coefficient depends on the shape of the object as well as the Reynolds number of the flow. For a sphere at a Reynolds number of $10^4$, $C_\mathrm{d}=0.47$ \citep{BakerEtAl1983}. Lower Reynolds numbers yield $C_\mathrm{d}>1$ \citep[e.g.][]{Lautrup2011}. For a short cylinder the drag coefficient is approximately twice as high, $C_\mathrm{d}=1.15$ at a Reynolds number of $10^4$ \citep{BakerEtAl1983}. These numbers are for solid bodies, so we expect variations between the literature numbers and our simulations. The effective cross section is initially the projected area of the cloud, i.e. $A=\pi r_\mathrm{c}^2$. Since all setups have the same cloud radius and $M_\mathrm{c}\propto\rho_\mathrm{c}$ we expect the acceleration to scale as $a_\mathrm{c}\propto\rho_\mathrm{c}^{-1}$ assuming that the effective area and the drag coefficient remain constant.

In a magnetized wind the cloud evolution differs. From previous studies several important effects have been identified. These include the changes of the growth rate of the Kelvin-Helmhotz instability \citep[e.g.][]{Chandrasekhar1961} and an overall enhancement or reduction of the rate at which the cloud is mixed into the wind depending on the magnetic field orientation \citep[e.g.][]{McCourtEtAl2015, CottleEtAl2020}. Additionally, the magnetic field can increase the transport of momentum from wind to cloud, such that the cloud is more effectively accelerated by the wind. This effect, known as magnetic draping \citep[e.g.][]{DursiPfrommer2008, McCourtEtAl2015}, has an additional contribution to the drag force, which reads
\begin{equation}
F_\mathrm{d} = \frac{1}{2}\rho_\mathrm{w}\left(v_\mathrm{w}^2+v_\mathrm{A}^2\right)C_\mathrm{d}A.
\end{equation}
The draping effect is strongest in the case of $\vektor{B}\perp \vektor{v}_\mathrm{w}$, where the magnetic field keeps the effective cross section at a maximum. The analytic estimates for $a_\mathrm{c}$ using $A=\pi r_\mathrm{c}^2$ and $C_\mathrm{d}=1$ are $a_c=0.82$, $0.16$ and $0.08\,\kmpers\,\mathrm{Myr}^{-1}$.

In order to compute an \emph{effective acceleration} for the entire cloud, we investigate the location and velocity of the integrated mass along the direction of the wind. We compute profiles along the $x$ coordinate and measure the position of the 25 percentile of the integrated mass, i.e. we define the position of the cloud to be located at the 25 percentile of the mass in the box along the $x$ direction. We compare several alternative numbers in Appendix ~\ref{sec:cloud-position}. The velocity of the 25 percent mark is shown in Fig.~\ref{fig:velocity-25percent-mark}. All simulations show a linear increase of the velocity over time. For the clouds with the lowest density, the cloud is destroyed after $\sim20-25,\,\mathrm{Myr}$. Once a noticeable fraction of the gas leaves the right-hand $x$ boundary the velocity profiles strongly fluctuate. We therefore only plot the values up to the point of cloud destruction for $n_\mathrm{c}=0.1\,\mathrm{cm}^{-3}$. To get the effective acceleration, we fit a linear function, $v(t) = a_\mathrm{eff} t$, to the velocities. We also compute an effective acceleration time $t_\mathrm{acc}=v_\mathrm{w}/a_\mathrm{eff}$. The values for all simulations are shown in Fig.~\ref{fig:effective-accelerations} together with the analytic estimates discussed above. The top panel shows the accelerations, the bottom one the acceleration time scale in units of the cloud crushing time. We apply a small offset in the abscissa in order to better distinguish between the run with and without self-gravity. We note a clear trend of the more massive clouds to experience a lower effective acceleration as expected. The theoretically estimated values are overall lower except for the most massive clouds. At a given cloud density the perpendicular magnetic field configuration accelerates the gas more efficiently than a non-magnetic wind. A wind with a field parallel to the flow is least efficient in speeding up the gas. With values of order $\sim0.2\,\mathrm{km\,s^{-1}\,Myr^{-1}}$ the dense clouds need $50\,\mathrm{Myr}$ to be sped up to $10\,\kmpers$. With values of $a_\mathrm{eff}\sim0.5-0.7\,\mathrm{km\,s^{-1}\,Myr^{-1}}$ a time of $20\,\mathrm{Myr}$ is required. The low density clouds reach this velocity after only a few Myr. In the case of $\Bperpv$ the acceleration follows the theoretical scaling with $a_\mathrm{eff}\propto n^{-1}$. This is most likely connected to the strong draping effect, which keeps the effective cross section of the cloud at a maximum, independent of the degree of fragmentation over time. The two other wind configurations result in smaller values for $a_\mathrm{eff}$ and also a different scaling. The fragmentation and partial dispersal of the cloud changes the effective cross section, the drag coefficient as well as the distribution of mass along the $x$-axis. At this point we refrain from disentangling the degeneracy between changes in all three quantities over time.

\section{Turbulence in the cloud}
\label{sec:turbulence}

\begin{figure}
\centering
\includegraphics[width=8cm]{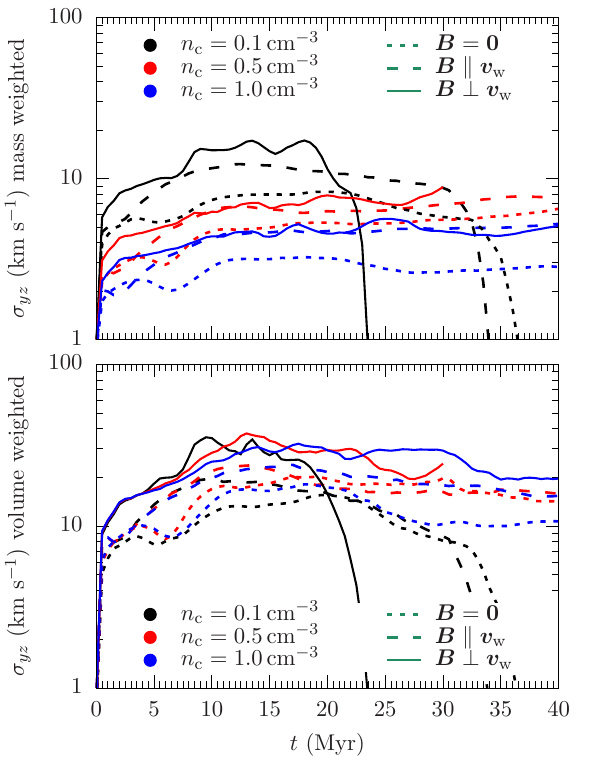}
\caption{Mass weighted (top) and volume weigthed (bottom) velocity dispersion in the $yz$ direction for all runs including self-gravity. This velocity dispersion is a simple measure for the turbulence in the cloud driven by the wind. The initial velocity dispersion in the cloud was set to only $1\,\kmpers$, so most of the measured motions are wind-driven, reaching up to 30 per cent of the wind velocity.}
\label{fig:yz-velocity-dispersion}
\end{figure}

Besides the net acceleration of the cloud, the wind also triggers motions perpendicular to the flow direction. We quantify the turbulence that is driven by the wind as the mass and volume weighted velocity dispersion of the $yz$ components of the velocity vector,
\begin{equation}
\sigma_{yz} = \left(\sum_i w_i\right)^{-1}\,\sum_i w_i\left(v_{i,y}^2 + v_{i,z}^2\right)^{1/2},
\end{equation}
where we sum over all cells $i$ in the computational domain and use the cell mass ($w_i = m_i$) and cell volume ($w_i=V_i$) for the weighted cell velocities ($v_i$). We restrict the analysis to the $yz$-component of the flow to avoid the complications in disentangling the net wind flow from turbulent motions in $x$. We show the time evolution of the velocity dispersion in Fig.~\ref{fig:yz-velocity-dispersion} for mass (top) and volume weighting (bottom). The wind-induced internal motions exceed the initial turbulence with a root-mean-square velocity of $1\,\kmpers$, so the stirring of the cloud is enhanced by the wind. The difference between the runs with and without self-gravity is negligible (not shown), so we conclude that most of the motions are actually driven by the wind rather than by gravitational contraction. The mass weighted values span a range from $2-3\,\kmpers$ for $n_\mathrm{c}=1\,\mathrm{cm}^{-3}$ up to $\sim10\,\kmpers$ for $n_\mathrm{c}=0.1\,\mathrm{cm}^{-3}$. The non-magnetic wind is least efficient in driving turbulence. Magnetised winds increase $\sigma_{yz}$ by a few $\kmpers$. The orientation of the magnetic field plays a minor role. In the mass weighted velocity dispersion the differences are insignificant. The volume weighted quantities are overall higher as expected with values ranging from $\sim10-20\,\kmpers$. The relative differences between the individual setups are smaller than in the mass weighted counterparts. The impact on the volume weighted velocity dispersion increases from non-magnetic, to parallel, to perpendicularly magnetised winds. With $\sigma_{yz}\sim10-20\,\kmpers$ the velocity dispersion perpendicular to the direction of the wind is approximately $10-20$ per cent of the wind speed. These numbers are broadly consistent with the analysis in \citep{BandaBarraganEtAl2016, BandaBarraganEtAl2018}. However, their numbers range from $\sim2-30$ per cent and show stronger temporal variations.

\section{Density evolution and chemical composition}
\label{sec:chemistry}

\begin{figure*}
\begin{minipage}{\textwidth}
\includegraphics[width=\textwidth]{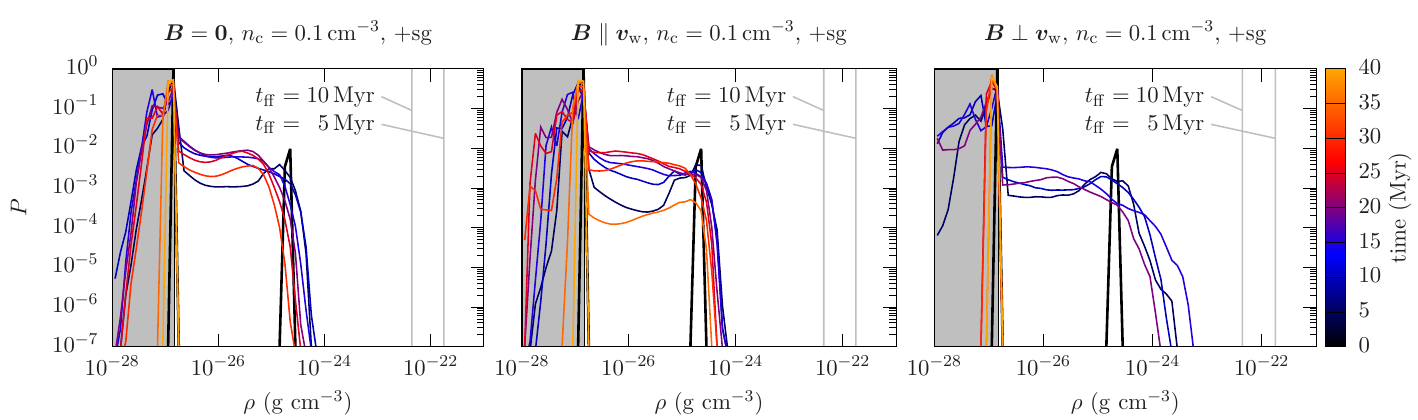}\\
\includegraphics[width=\textwidth]{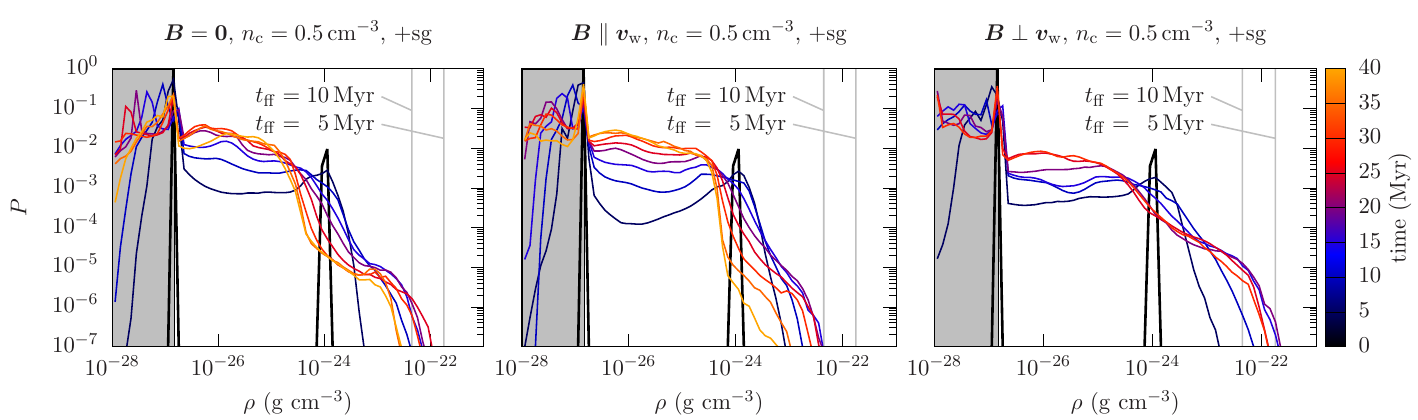}\\
\includegraphics[width=\textwidth]{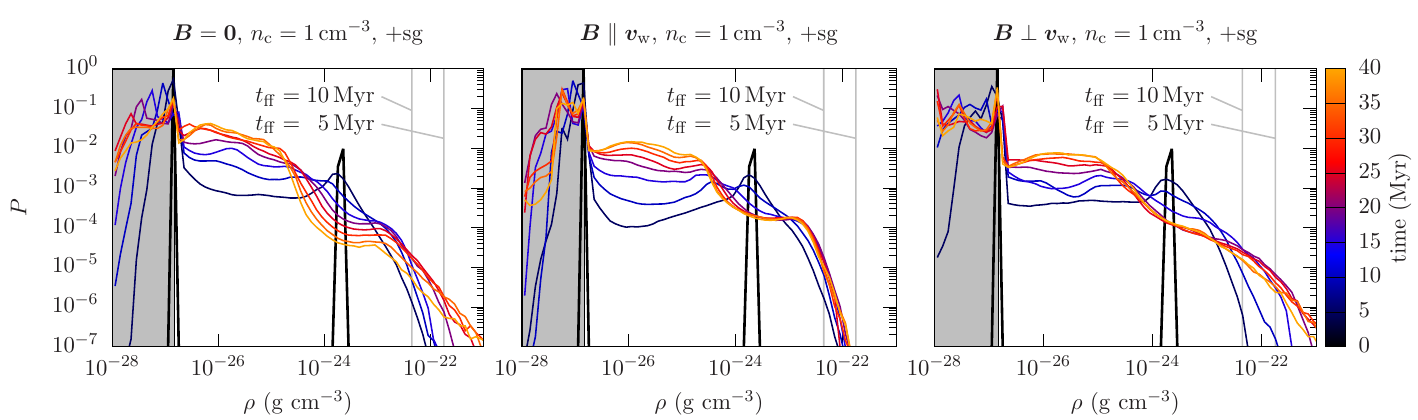}
\caption{Time evolution of the density PDF for all simulations including self-gravity. From top to bottom we increase the cloud density. From left to right we alter the wind properties. The time is colour-coded. The grey area indicates densities lower than the incoming wind, i.e. volumes at those densities form due to the turbulent interactions. The two vertical lines mark the densities corresponding to a free-fall time of $5$ and $10\,\mathrm{Myr}$, which is 12.5 and 25 per cent of the total simulation time. Magnetized winds with a field perpendicular to the flow are most efficient at compressing the gas, which is reflected in the high-density tail. Clouds with $n_c=0.5\percc$ reach densities at which the gas could gravitationally collapse within the simulation time. Clouds with an initial density of $1\,\mathrm{cm^{-3}}$ form a high-density tail that is consistent with a gravitationally driven tail \citep[e.g.][]{GirichidisEtAl2014}.}
\label{fig:density-pdf-time-evol}
\end{minipage}
\end{figure*}

\subsection{Density evolution}

We follow the evolution of the cloud quantitatively using the probability density function (PDF) of the total gas density. Fig.~\ref{fig:density-pdf-time-evol} illustrates the time evolution for all simulations including self-gravity. From top to bottom we show the three different cloud densities, from left to right we vary the magnetic field properties in the wind. The vertical lines indicate the densities corresponding to a free-fall time of 5 and 10\,Myr. The thick black lines depict the density distribution of the initial conditions. The low density clouds (top row) do not show a strong compression of gas before the destruction of the cloud. An exception is the cloud exposed to a wind with a perpendicular magnetic field with a temporary maximum compression of about one order of magnitude above the initial conditions. In the other two cases (i.e. parallel or no magnetic field) the high density part of the PDF remains close to the initial values before the cloud is quickly pushed out of the box. The medium density clouds ($n_\mathrm{c}=0.5\,\percc$, middle row) show the evolution of a high-density tail, up to two orders of magnitude above the initial maximum density. In particular, for the non-magnetic wind and the perpendicular field orientation, the high density tail remains until the end of the simulation and forms molecular hydrogen. For the $\Bparv$ runs the cloud eventually disperses with little gas above the initial density. In all high density clouds ($n_\mathrm{c}=1\,\percc$, bottom row) the highest density due to interaction with the wind and self-gravity exceeds the initial values by at least two orders of magnitude. Again, the non-magnetized wind and the wind with a perpendicular magnetic field most efficiently compress the gas. In both cases the density reaches peaks, which correspond to free-fall times of about a Myr.

\subsection{Evolution of atomic and molecular hydrogen}

\begin{figure}
\includegraphics[width=8cm]{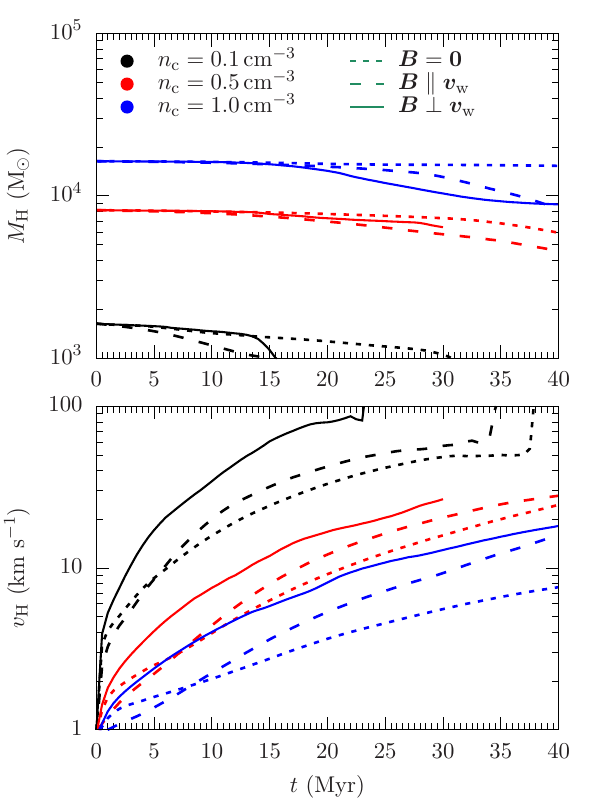}
\caption{Time evolution of the atomic gas fraction (top panel) as well as the corresponding mass weighted velocity in $x$-direction (bottom panel). Shown are all simulations including self-gravity. In all cases the atomic gas mass decreases over time. For the medium density clouds the lost atomic gas is heated and dispersed by the wind. For the high density clouds the lost atomic gas is converted to H$_2$. The velocities are systematically increasing from non-magnetic, to $\Bparv$, to $\Bperpv$ winds.}
\label{fig:time-evol-Ha-mass-velo}
\end{figure}

\begin{figure*}
\begin{minipage}{\textwidth}
\includegraphics[width=\textwidth]{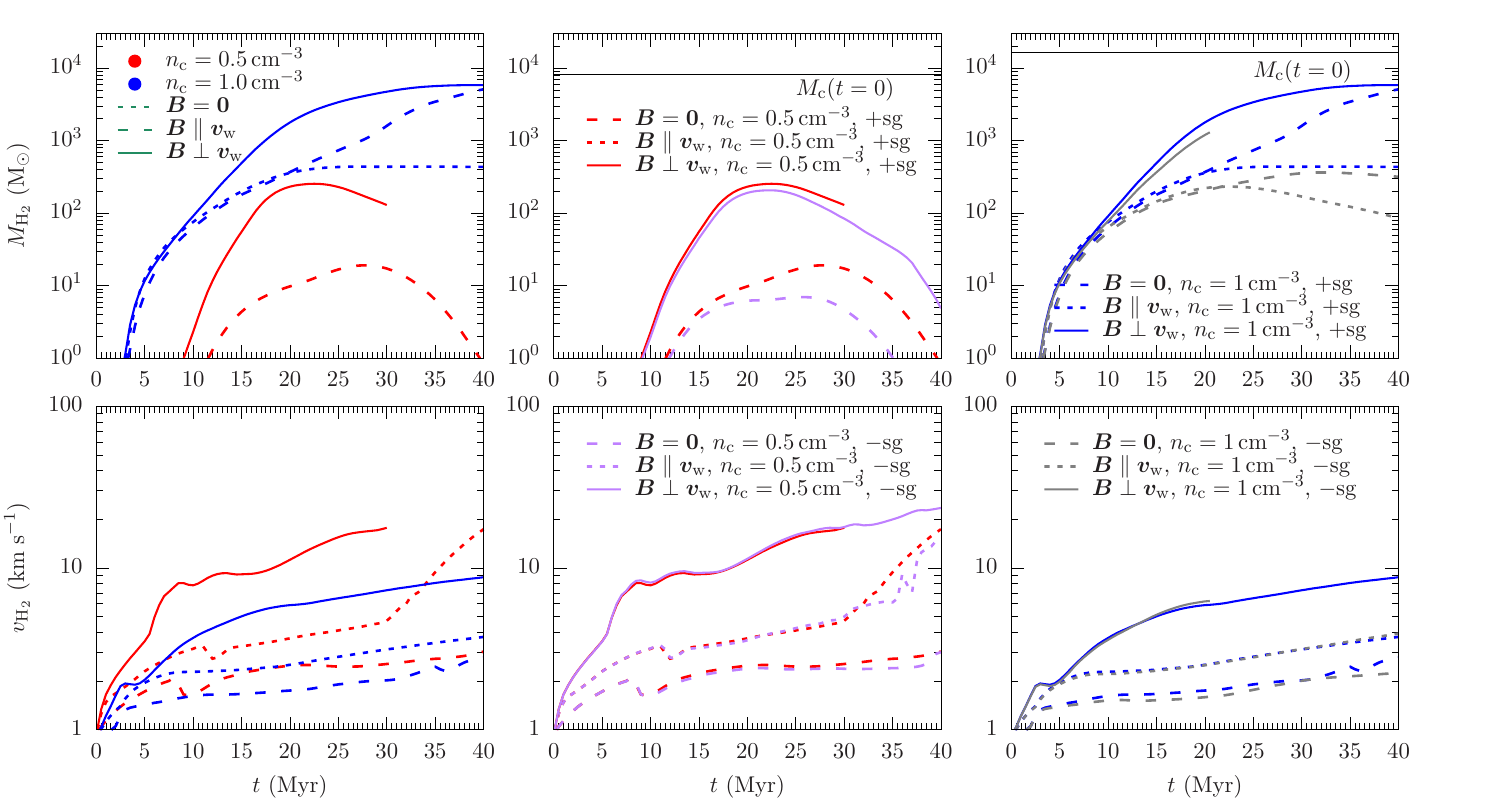}
\caption{Time evolution of the molecular gas mass (top panels) as well as the molecular mass weighted velocity in the $x$-direction (bottom panels). The left-hand panel shows all simulations including self-gravity. The central and right-hand panel compare the effect of self-gravity for a the medium ($n_\mathrm{c}=0.5\,\percc$) and high density ($n_\mathrm{c}=1\,\percc$) cloud, respectively. The medium density clouds form H$_2$ in the first half of the simulated time. The wind configuration has an important impact on how much H$_2$ forms. During the second half of the time the molecular gas is destroyed again. For the dense clouds the wind configuration and self-gravity are equally important. Here a stable fraction of molecular gas can survive until the end.}
\label{fig:time-evol-H2-mass-velo}
\end{minipage}
\end{figure*}

We show the evolution of the atomic hydrogen for all simulations including self-gravity in Fig.~\ref{fig:time-evol-Ha-mass-velo}. The top panel shows the mass, the bottom one the atomic mass weighted velocity over time. We note the systematic effect of the density and the magnetic field in the velocity profiles. From high to low density the effective velocity increases, the same holds for non-magnetic, to $\Bparv$, $\Bperpv$ winds. The low density clouds reach velocities of approximately $40-60$ per cent of the wind speed before they are dissolved. The medium and high density clouds only reach between $5$ and $30$ per cent of $v_\mathrm{w}$, but the velocities are still increasing at the end of the simulation. All clouds lose atomic gas mass over time. In the case of $n_\mathrm{c}=0.5\,\percc$ a noticeable fraction of this atomic loss is converted into hot ionized gas. In the case of $n_\mathrm{c}=1.0\,\percc$ most of the lost gas is converted to the molecular phase as we will discuss in more detail below.

Only the medium and high density clouds form a significant fraction of molecular gas. Fig.~\ref{fig:time-evol-H2-mass-velo} depicts the total mass in molecular form (top panels) and the velocity, weighted with the molecular gas mass (bottom panels). In the left-hand panel we show all simulations including self-gravity. The central panel illustrates the differences caused by self-gravity for the medium density clouds ($n_\mathrm{c}=0.5\,\mathrm{cm^{-3}}$). The right-hand panel shows the same for the clouds with $n_\mathrm{c}=1\,\mathrm{cm^{-3}}$. All setups with a low cloud density only form tiny fractions of molecular gas (not shown). For the medium density clouds molecular gas forms after approximately 10\,Myr (red lines). The properties of the wind have a noticeable impact on how much H$_2$ forms. Whereas the perpendicular field allows for the temporal formation of $200\,\mathrm{M}_\odot$ of H$_2$, the non-magnetic wind only leads to an order of magnitude less molecular gas. The wind with a parallel field further suppresses the condensation into a molecular phase. The difference between runs with and without self-gravity is a secondary effect. In all cases with $n_\mathrm{c}=0.5\,\mathrm{cm^{-3}}$ the molecular gas is destroyed towards the end of the simulation. Significant fractions of the total mass can only be converted to the molecular phase in the high density clouds (right-hand panels). Again, the magnetic field has an important impact on the evolution, but for the comparably high densities, the impact of self-gravity is similarly important. Whether self-gravity is included or not mainly changes the evolution during the second half of the simulation. When the field is parallel to the wind, the molecular gas fraction decreases again after $t\sim25\,\mathrm{Myr}$. Including self-gravity, a total amount of $300\,\mathrm{M}_\odot$ of H$_2$ ($2-3$ per cent of the total mass) remains until the end of the simulation. The non-magnetic wind allows for more compression perpendicular to the wind such that even without self-gravity a stable fraction of $2-3$ per cent of the hydrogen survives in molecular form. The additional contraction provided by self-gravity can increase that fraction by an order of magnitude up to the end of the simulation. Again, the most efficient way of compressing the gas and forming molecular hydrogen is the perpendicular field configuration. In this case the inclusion of self-gravity has a minor effect up to the end of the simulated time. For the velocities there is a clear trend that winds with a perpendicular field accelerate the molecular gas most efficiently up to velocities of $\sim10-20\,\kmpers$ until the end of the simulation.

\begin{figure*}
\begin{minipage}{0.8\textwidth}
\includegraphics[width=0.9\textwidth]{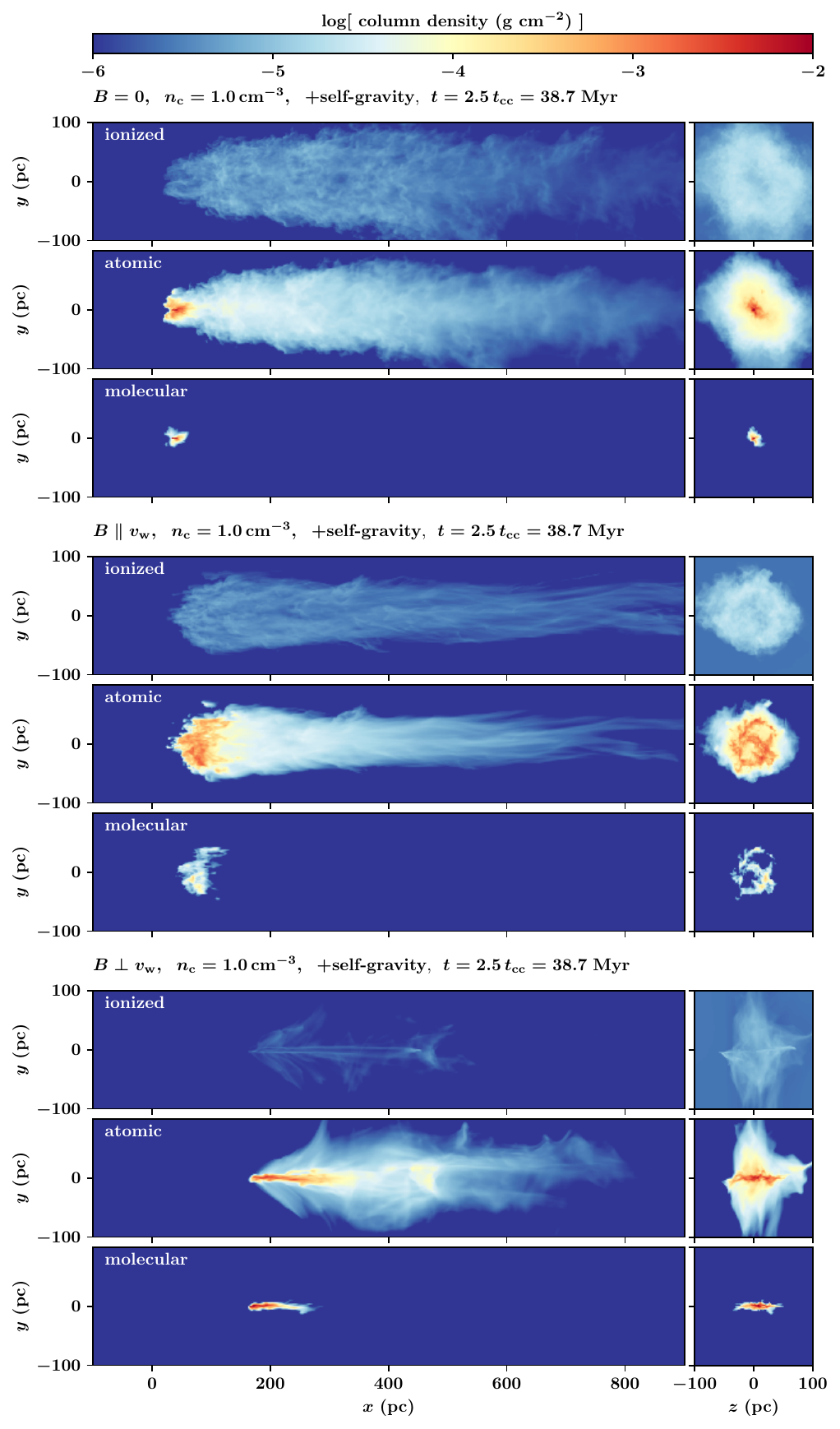}
\caption{Chemical composition for all simulations with $n_\mathrm{c}=1\,\percc$ and self-gravity after $2.5\,t_\mathrm{cc}$. The non-magnetic wind triggers the formation of a spherical molecular cloud with a broad tail composed of atomic and ionized gas. A parallel field in the wind causes molecular fragments distributed over a larger area. The thinner tail contains again a mix of atomic and ionized gas. The perpendicular field can compress most of the gas into a coherent elongated molecular cloud. Almost all the gas is either molecular or atomic, there are only spurs of ionized gas in the tail.}
\label{fig:cd-chem-t40-L6-B0-d100-ng}
\end{minipage}
\end{figure*}

Fig.~\ref{fig:cd-chem-t40-L6-B0-d100-ng} illustrates the chemical composition in the box for the high density clouds including self-gravity and the three different wind configurations. In the presence of a non-magnetic wind, a turbulent atomic stream emerges from the cloud, whose central region condenses to form a compact molecular core close to the initial location of the cloud. In the case of a magnetized wind with a parallel field the molecular gas is distributed into numerous individual fragments. By the end of the simulation the molecular gas can travel a distance of approximately $100\,\mathrm{pc}$ from the original position of the cloud. The atomic tail simply follows the magnetic field lines and shows a coherent stream with little local vertical turbulent perturbations, see also Fig.~\ref{fig:dens-slice-mag-strm}. The perpendicular field does not only push the molecular gas out to $x\sim200\,\mathrm{pc}$, but also forms an elongated molecular stream along the flow rather than a spherical cloud. The atomic tail shows substructures and perturbations that are larger than in the other two simulations. There is almost no over-density of ionized gas in the tail. 

\subsection{Wind-triggered H$_2$ formation}

\begin{figure*}
\begin{minipage}{\textwidth}
\includegraphics[width=\textwidth]{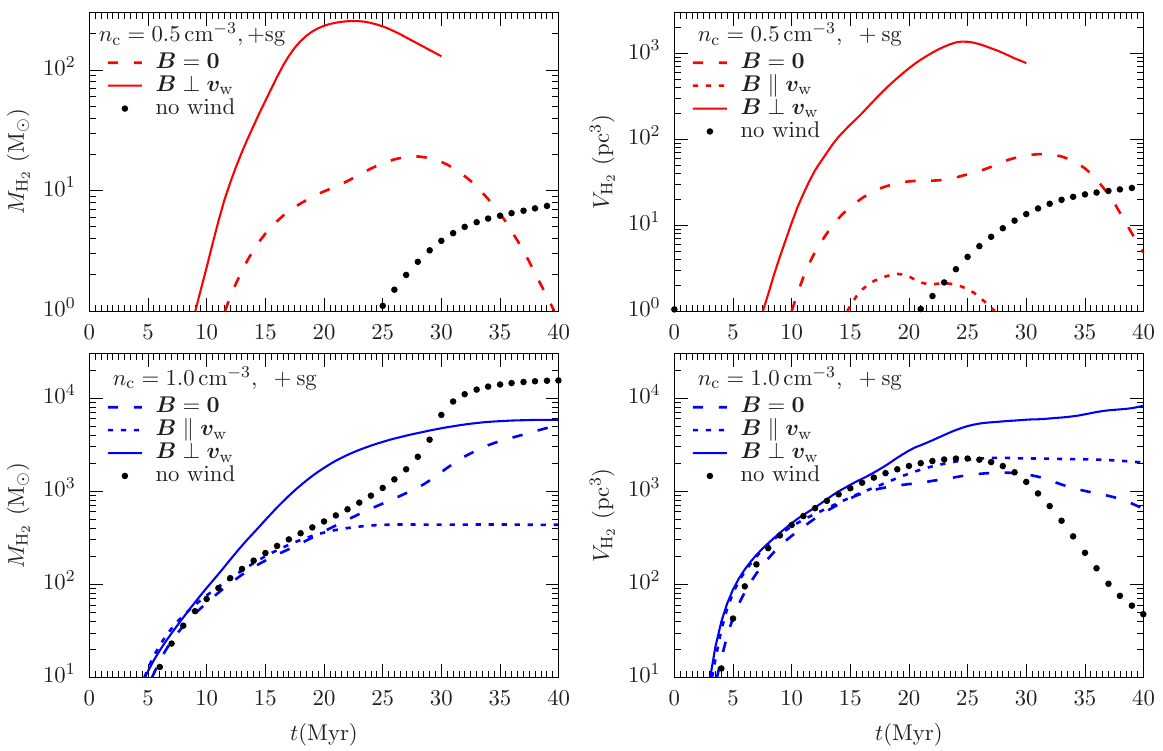}
\caption{Comparison of the H$_2$ formation over time for clouds with an initial number density of $n_\mathrm{c}=0.5\,\mathrm{cm}^{-3}$ (top panels) and $n_\mathrm{c}=1\,\mathrm{cm}^{-3}$ (bottom panels). We show the total mass in H$_2$ (left-hand panels) as well as the H$_2$ mass weighted volume (right-hand panels). For the medium density clouds ($n_\mathrm{c}=0.5\,\percc$) the wind both speeds up the formation of molecular gas and increases the peak of the H$_2$ mass. For the dense clouds ($n_\mathrm{c}=1\,\percc$) the perpendicular field configuration ($\vektor{B}\perp\vektor{v}_\mathrm{w}$) can speed up the formation of H$_2$ for most of the simulation time. After $t\sim30\,\mathrm{Myr}$ the isolated cloud collapses and confined all the mass in a small volume.}
\label{fig:H2-formation-comparison}
\end{minipage}
\end{figure*}

\begin{figure}
\includegraphics[width=8cm]{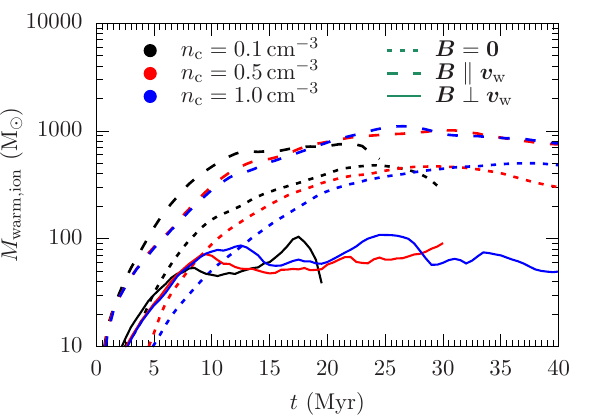}
\caption{Time evolution of the warm ionized gas. The total mass mainly depends on the the wind properties rathre than the cloud mass. We only show the low density up to the point of their destruction and their transport through the right-hand boundary. We find the lowest mass in $\vektor{B}\perp\vektor{v}_\mathrm{w}$, followed by the non-magnetic wind. For $\vektor{B}\parallel\vektor{v}_\mathrm{w}$ the warm ionized mass is largest.}
\label{fig:time-evol-WIM}
\end{figure}

We address the question of how much a wind can speed up or trigger the formation of molecular gas by investigating the total mass in molecular hydrogen as well as the volume that is occupied by H$_2$. The analysis of the mass is a simple, obvious measure. However, at late times we need to take into account that a cloud which is not exposed to a wind can continue to cool and collapse and in principle condense all the gas in molecular form in a single computational cell. Contrary, the wind will remove some material. We thus also investigate how much volume the molecular phase occupies,
\begin{equation}
V_\mathrm{H_2} = \sum_i f_\mathrm{H_2,i} V_i
\end{equation}
with the fractional abundance of H$_2$, $f_\mathrm{H_2,i}$, and the volume, $V_i$, in cell $i$ and the sum includes all cells of the simulation domain.

Fig.~\ref{fig:H2-formation-comparison} shows the temporal evolution of H$_2$ for medium (top) and high density clouds (bottom) for both the mass (left-hand panels) and the volume (right-hand panels). For the medium density clouds the wind can accelerate the formation of H$_2$ and also increase the peak mass and occupied volume except for the parallel field configuration, $\vektor{B}\parallel\vektor{v}_\mathrm{w}$. For $\vektor{B}\perp\vektor{v}_\mathrm{w}$ molecular gas forms 15\,Myr earlier than in the no-wind case and the peak mass in H$_2$ is more than an order of magnitude larger. The volumes behave similarly. In this case we can speak of wind-triggered H$_2$ formation. For the dense clouds ($n_\mathrm{c}=1\,\percc$, bottom panels) the initial evolution is very similar for all setups except $\vektor{B}\perp\vektor{v}_\mathrm{w}$, where $M_\mathrm{H_2}$ increases more rapidly. After $t\sim30\,\mathrm{Myr}\approx2t_\mathrm{cc}$, the isolated cloud collapses and converts basically all the mass into molecular form into a very small volume. This is reflected in the volume plot in the right-hand panel, see also the column density plots in Fig~\ref{fig:coldens-noW}. The clouds exposed to a wind confine the molecular gas in a significantly larger volume, i.e. the wind triggers a more distributed H$_2$ formation.

\subsection{Warm ionized gas}

Finally, we measure the warm ionized medium (WIM) with temperatures between 8000 and $3\times10^5\,\mathrm{K}$ and an ionisation fraction larger than 0.5. Initially, there is only warm atomic and hot ionized gas, so all the warm ionized gas is solely a result of the interaction of the cloud with the wind. Fig.~\ref{fig:time-evol-WIM} shows the time evolution of the warm ionized gas for all simulations with self-gravity. We note that the main parameter that changes the total amount of the WIM is not the density of the cloud but mainly the magnetic field properties in the wind. All winds with a perpendicular field keep the clouds mostly intact due to the magnetic draping effect and reduce the mixing of warm atomic and hot ionized gas at the cloud wind interface. The resulting mass in the WIM converges after about 10\,Myr of evolution to give $50-100\,\mathrm{M}_\odot$. For the non-magnetic wind the masses converge after $\sim25\,\mathrm{Myr}$ at a value of $\sim500\,\mathrm{M}_\odot$. The highest masses in the WIM are reached for $\vektor{B}\parallel\vektor{v}_\mathrm{w}$ with about $1000\,\mathrm{M}_\odot$. We highlight that this is 60 per cent of the initial mass of the low density cloud. The parallel field configuration delays or completely suppresses the contraction perpendicular to the magnetic field and thereby maximizes the mixing that leads to the WIM.

\section{Discussion and Conclusions}
\label{sec:discussion-conclusion}

\subsection{Cloud survival}
For clouds exposed to a wind the question arises whether the clouds are fed by condensation of wind material and grow, whether they are effectively unaltered or whether the wind eventually disrupts the cloud. In absence of stabilizing mechanisms the clouds are likely to be destroyed by the efficient growth of the Rayleigh-Taylor instability and the shear-driven Kelvin-Helmhotz instability \citep{CooperEtAl2009, ScannapiecoBrueggen2015, SchneiderRobertson2015, SchneiderRobertson2017}. Stabilizing mechanisms can be the cooling and a resulting contraction that reduces the effective interaction cross section with the wind. It could also be the magnetic field configuration, which suppresses instabilities or -- mostly for dense clouds -- self-gravity, which leads to faster contraction and even collapse. Cooling can lead to fragmentation of the clouds \citep{McCourtEtAl2018}, but can also help to stabilize them by suppressing the mixing \citep{SparrePfrommerVogelsberger2019}. Stabilized condensations can grow in mass due to newly accreted gas from the wind \citep{MarinacciEtAl2010, ArmillottaFraternaliMarinacci2016,ArmillottaEtAl2017,GronkeOh2018, GronkeOh2020a}. Magnetic fields help to keep the gas confined and increase the acceleration efficiency \citep{McCourtEtAl2015, BandaBarraganEtAl2016, BandaBarraganEtAl2018, BandaBarraganEtAl2019, GronnowEtAl2018, SparrePfrommerEhlert2020}. All of the above mentioned simulations have focused on overall lower densities and higher temperatures compared to this study. However, the surviving fragments develop strong density fluctuations, which are comparable to the initial configurations in our models.

We evolve the systems for a time scale of a few cloud crushing times, which is sufficient to examine the formation and partial destruction of molecular hydrogen. For medium and high density clouds, the strong compression of the gas at the interface between the wind and the cloud is sufficient to trigger significant H$_2$ formation. In the molecular regions self-gravity can be important to determine the fate of the cloud \citep[see also][]{MurrayEtAl1993}. For the high density cloud with $n_\mathrm{c}=1\,\mathrm{cm}^{-3}$ the Jeans length quickly shortens to values below the clump structures visible in the column density plots. Here the magnetic field plays an important role. With a non-magnetic wind ($\vektor{B}=\vektor{0}$) and in the case of $\vektor{B}\perp \vektor{v}_\mathrm{w}$, 30 per cent of the mass is located in regions, in which the Jeans length is smaller than $10\,\mathrm{pc}$. For the magnetized wind with $\vektor{B} \parallel \vektor{v}_\mathrm{w}$ this fraction of mass is only about 0.3 per cent of the total gas mass. For the former two runs, we thus expect triggered collapse, a surviving fraction of molecular gas and possibly triggered star formation in the clouds.

\subsection{Observations}

Cold and molecular outflows are a common phenomenon in galaxies \citep[see e.g. recent reviews by][]{Rupke2018, VeilleuxEtAl2020}. On the one hand it is plausible that the observed cold outflows have undergone the transition to the molecular phase in the dense regions of the galactic disc and have been pushed out as an outflow. On the other hand, numerous compact high-velocity clouds have been observed in atomic hydrogen with velocities that are consistent with outflow motions \citep{McClureGriffiths2013, DiTeodoroEtAl2018,LockmanDiTeodoroMcClureGriffiths2020}. These clouds are observed with comparable properties to our simulated systems ($r\sim10-40\,\mathrm{pc}$, $M_\mathrm{H}\sim10-10^5\,\mathrm{M}_\odot$). Based on our simulations it seems plausible that the molecular gas forms in these high-velocity clouds high above the galactic disc. The observed velocities of these clouds relative to the Milky Way ranges from $\sim50-200\,\kmpers$ \citep[][and references therein]{VeilleuxEtAl2020} and is consistent with the numerical experiments presented here. The triggered collapse and the possible star formation at heights of more than a kpc above the disc is in agreement with our simulations.

\subsection{Conclusion}

We have performed magneto-hydrodynamical simulations of a cloud exposed to a hot wind with a velocity of $100\,\kmpers$. We vary the density contrast between the cloud and the ambient medium over one order of magnitude. We model three different types of winds, namely a non-magnetized wind and two magnetized counterparts with the magnetic field oriented parallel ($\Bparv$) or perpendicular to the wind direction ($\Bperpv$). We specifically focus on the thermal state of the gas down to temperatures of $\sim10\,\mathrm{K}$ and include the formation of molecular hydrogen together with the shielding of the gas from interstellar radiation. We also investigate the impact of self-gravity as a stabilizer for the dense condensations and an accelerator of the formation of molecular gas.

Of the parameters that we vary, the density has the biggest impact on the system evolution. In scale-free setups, e.g. without cooling, this simply relates to the density contrast but since our models are not scale-free the actual density is important. We can summarize our results as follows.

\begin{itemize}
\item The low density ($n_\mathrm{c}=0.1\,\percc$) clouds are dissolved within a few cloud crushing times, irrespective of the magnetic field configuration and self-gravity, so all the gas is quickly accelerated. The compression of the gas is not strong enough to form any molecular hydrogen.
\item The dynamics of medium density clouds ($n_\mathrm{c}=0.5\,\percc$) is perceptibly affected by the wind details. Only the wind with $\Bperpv$ can accelerate the cloud ($10-20$ percent of wind speed within $2-3\,t_\mathrm{cc}$) and form a noticeable fraction of molecular hydrogen (a few per cent of the initial cloud mass). The molecular hydrogen forms over a time scale of $\sim20\,\mathrm{Myr}$ ($\sim2\,t_\mathrm{cc}$) but is not stabilized by self-gravity, so the continuous wind also destroys most of it at later times. 
\item All clouds with high initial density ($n_\mathrm{c}=1\,\percc$) form significant fractions of molecular hydrogen. The non-magnetic and perpendicular ($\Bperpv$) wind configuration can convert more than $10-20$ per cent of the total cloud mass into H$_2$. This fraction is stable or even increases over time, so the dense condensations are likely to remain confined and collapse to form stars. A wind with a magnetic field parallel to the flow still triggers H$_2$ formation, but the contraction perpendicular to the field is suppressed, so the cloud fragments into smaller molecular cloudlets. Without self-gravity, those H$_2$ cloudlets do not survive over time in this wind configuration.
\item Comparing simulations with and without winds reveals that the wind enhances and accelerates H$_2$ formation for our setups with medium and high density clouds ($0.5$ and $1\,\percc$.) In these cases a hot wind can trigger the formation of molecular gas, which can become gravitationally unstable within a time scale of less than a cloud crushing time. The low density clouds with $0.1\,\percc$ cannot be compressed enough to trigger the formation of H$_2$.
\item In all cases, in which molecular gas forms, the densities are high enough and the dense masses are compact enough to trigger collapse and lead to star formation. This suggests that -- if the initial (high velocity) clouds are present in the circum-galactic medium -- molecular gas could form in the outflow thus enabling star formation far away from the galactic disc.
\item The acceleration ($a_\mathrm{eff}$) of a cloud exposed to a $\Bperpv$ wind nicely follows the analytic scaling based on the drag force, $a_\mathrm{eff}\propto n^{-1}$. The actual values of the acceleration is a factor of a few lower than the estimates for an ideal, stable gas configuration of a sphere. The derived acceleration time scales range from $2-20\,t_\mathrm{cc}$ from low to high-density clouds. Non-magnetic and $\Bparv$ winds are less efficient in accelerating the gas by another factor of a few compared to the $\Bperpv$ wind case. The derived acceleration times scales are almost an order of magnitude larger, so $10-200\,t_\mathrm{cc}$.
\item The formation of warm ionized gas occurs due to mixing of the hot ionized wind with the warm atomic cloud. The total mass in this phase only depends on the wind configuration, i.e. is independent of the cloud density. Winds with a parallel magnetic field ($\Bperpv$) confine the gas via magnetic draping, reduce mixing and lead to the lowest mass in the WIM ($\sim50-100\,\mathrm{M}_\odot$). Non-magnetic winds produce up to $\sim500\,\mathrm{M}_\odot$. For $\Bparv$ the mixing is strongest and up to $\sim1000\,\mathrm{M}_\odot$ of warm ionized gas. For the low density clouds this corresponds to 60 per cent of the initial cloud mass.
\end{itemize}

\section*{Acknowledgements}
We thank the anonymous referee for clear comments that helped improving the manuscript. We thank Martin Sparre and Christoph Pfrommer for inspiring discussions.
PG and TB acknowledge funding from the European Research Council under ERC-CoG grant CRAGSMAN-646955.
TN acknowledges support from the Deutsche Forschungsgemeinschaft (DFG, German Research Foundation) under Germany’s Excellence Strategy - EXC-2094 - 390783311 from the DFG Cluster of Excellence "ORIGINS”.
SW gratefully acknowledges the European Research Council under the European Community’s Framework Programme FP8 via the ERC Starting Grant RADFEEDBACK (project number 679852). SW further thanks the Deutsche Forschungsgemeinschaft (DFG) for funding through SFB 956 \emph{The conditions and impact of star formation}, sub-project C5. SW also thanks the Bonn-Cologne-Graduate School.
The authors acknowledge computational resources from the Gauss Center for Supercomputing (http://www.gauss-centre.eu/) and the Leibniz-Rechenzentrum Garching (http://www.lrz.de) under project pn34ma. In addition we thank the Max Planck Computing and Data Facility (MPCDF) for computing time and data storage.
The software used in this work was developed in part by the DOE NNSA ASC- and DOE Office of Science ASCR-supported Flash Center for Computational Science at the University of Chicago.

\section*{Data Availability}

The simulation data and data analysis scripts for this study will be shared upon request to the corresponding author.




\bibliographystyle{mnras}
\bibliography{girichidis.bib,astro.bib} 

\begin{thebibliography}{}
\makeatletter
\relax
\def\mn@urlcharsother{\let\do\@makeother \do\$\do\&\do\#\do\^\do\_\do\%\do\~}
\def\mn@doi{\begingroup\mn@urlcharsother \@ifnextchar [ {\mn@doi@}
  {\mn@doi@[]}}
\def\mn@doi@[#1]#2{\def\@tempa{#1}\ifx\@tempa\@empty \href
  {http://dx.doi.org/#2} {doi:#2}\else \href {http://dx.doi.org/#2} {#1}\fi
  \endgroup}
\def\mn@eprint#1#2{\mn@eprint@#1:#2::\@nil}
\def\mn@eprint@arXiv#1{\href {http://arxiv.org/abs/#1} {{\tt arXiv:#1}}}
\def\mn@eprint@dblp#1{\href {http://dblp.uni-trier.de/rec/bibtex/#1.xml}
  {dblp:#1}}
\def\mn@eprint@#1:#2:#3:#4\@nil{\def\@tempa {#1}\def\@tempb {#2}\def\@tempc
  {#3}\ifx \@tempc \@empty \let \@tempc \@tempb \let \@tempb \@tempa \fi \ifx
  \@tempb \@empty \def\@tempb {arXiv}\fi \@ifundefined
  {mn@eprint@\@tempb}{\@tempb:\@tempc}{\expandafter \expandafter \csname
  mn@eprint@\@tempb\endcsname \expandafter{\@tempc}}}

\bibitem[\protect\citeauthoryear{{Armillotta}, {Fraternali}  \&
  {Marinacci}}{{Armillotta} et~al.}{2016}]{ArmillottaFraternaliMarinacci2016}
{Armillotta} L.,  {Fraternali} F.,   {Marinacci} F.,  2016, \mn@doi [\mnras]
  {10.1093/mnras/stw1930}, \href
  {https://ui.adsabs.harvard.edu/abs/2016MNRAS.462.4157A} {462, 4157}

\bibitem[\protect\citeauthoryear{{Armillotta}, {Fraternali}, {Werk},
  {Prochaska}  \& {Marinacci}}{{Armillotta} et~al.}{2017}]{ArmillottaEtAl2017}
{Armillotta} L.,  {Fraternali} F.,  {Werk} J.~K.,  {Prochaska} J.~X.,
  {Marinacci} F.,  2017, \mn@doi [\mnras] {10.1093/mnras/stx1239}, \href
  {https://ui.adsabs.harvard.edu/abs/2017MNRAS.470..114A} {470, 114}

\bibitem[\protect\citeauthoryear{Baker, Cox, Westine, Kulesz  \&
  Strehlow}{Baker et~al.}{1983}]{BakerEtAl1983}
Baker W.~E.,  Cox P.~A.,  Westine P.~S.,  Kulesz J.~J.,   Strehlow R.~A.,
  1983, Explosion Hazards and Evaluation.
Elsevier, Amsterdam

\bibitem[\protect\citeauthoryear{{Bakes} \& {Tielens}}{{Bakes} \&
  {Tielens}}{1994}]{BakesTielens1994}
{Bakes} E.~L.~O.,  {Tielens} A.~G.~G.~M.,  1994, \mn@doi [\apj]
  {10.1086/174188}, \href {http://adsabs.harvard.edu/abs/1994ApJ...427..822B}
  {427, 822}

\bibitem[\protect\citeauthoryear{{Banda-Barrag{\'a}n}, {Parkin}, {Federrath},
  {Crocker}  \& {Bicknell}}{{Banda-Barrag{\'a}n}
  et~al.}{2016}]{BandaBarraganEtAl2016}
{Banda-Barrag{\'a}n} W.~E.,  {Parkin} E.~R.,  {Federrath} C.,  {Crocker} R.~M.,
    {Bicknell} G.~V.,  2016, \mn@doi [\mnras] {10.1093/mnras/stv2405}, \href
  {https://ui.adsabs.harvard.edu/abs/2016MNRAS.455.1309B} {455, 1309}

\bibitem[\protect\citeauthoryear{{Banda-Barrag{\'a}n}, {Federrath}, {Crocker}
  \& {Bicknell}}{{Banda-Barrag{\'a}n} et~al.}{2018}]{BandaBarraganEtAl2018}
{Banda-Barrag{\'a}n} W.~E.,  {Federrath} C.,  {Crocker} R.~M.,   {Bicknell}
  G.~V.,  2018, \mn@doi [\mnras] {10.1093/mnras/stx2541}, \href
  {https://ui.adsabs.harvard.edu/abs/2018MNRAS.473.3454B} {473, 3454}

\bibitem[\protect\citeauthoryear{{Banda-Barrag{\'a}n}, {Zertuche}, {Federrath},
  {Garc{\'\i}a Del Valle}, {Br{\"u}ggen}  \& {Wagner}}{{Banda-Barrag{\'a}n}
  et~al.}{2019}]{BandaBarraganEtAl2019}
{Banda-Barrag{\'a}n} W.~E.,  {Zertuche} F.~J.,  {Federrath} C.,  {Garc{\'\i}a
  Del Valle} J.,  {Br{\"u}ggen} M.,   {Wagner} A.~Y.,  2019, \mn@doi [\mnras]
  {10.1093/mnras/stz1040}, \href
  {https://ui.adsabs.harvard.edu/abs/2019MNRAS.486.4526B} {486, 4526}

\bibitem[\protect\citeauthoryear{{Banda-Barrag{\'a}n}, {Br{\"u}ggen}, {Heesen},
  {Scannapieco}, {Cottle}, {Federrath}  \& {Wagner}}{{Banda-Barrag{\'a}n}
  et~al.}{2020}]{BandaBarraganEtAl2021}
{Banda-Barrag{\'a}n} W.,  {Br{\"u}ggen} M.,  {Heesen} V.,  {Scannapieco} E.,
  {Cottle} J.,  {Federrath} C.,   {Wagner} A.~Y.,  2020, arXiv e-prints, \href
  {https://ui.adsabs.harvard.edu/abs/2020arXiv201105240B} {p. arXiv:2011.05240}

\bibitem[\protect\citeauthoryear{{Beck}}{{Beck}}{2015}]{Beck2015}
{Beck} R.,  2015, \mn@doi [\aapr] {10.1007/s00159-015-0084-4}, \href
  {https://ui.adsabs.harvard.edu/abs/2015A&ARv..24....4B} {24, 4}

\bibitem[\protect\citeauthoryear{{Bellomi}, {Godard}, {Hennebelle}, {Valdivia},
  {Pineau des For{\^e}ts}, {Lesaffre}  \& {P{\'e}rault}}{{Bellomi}
  et~al.}{2020}]{BellomiEtAl2020}
{Bellomi} E.,  {Godard} B.,  {Hennebelle} P.,  {Valdivia} V.,  {Pineau des
  For{\^e}ts} G.,  {Lesaffre} P.,   {P{\'e}rault} M.,  2020, \mn@doi [\aap]
  {10.1051/0004-6361/202038593}, \href
  {https://ui.adsabs.harvard.edu/abs/2020A&A...643A..36B} {643, A36}

\bibitem[\protect\citeauthoryear{{Bergin}, {Hartmann}, {Raymond}  \&
  {Ballesteros-Paredes}}{{Bergin} et~al.}{2004}]{Bergin2004}
{Bergin} E.~A.,  {Hartmann} L.~W.,  {Raymond} J.~C.,   {Ballesteros-Paredes}
  J.,  2004, \mn@doi [\apj] {10.1086/422578}, \href
  {http://cdsads.u-strasbg.fr/abs/2004ApJ...612..921B} {612, 921}

\bibitem[\protect\citeauthoryear{{Bigiel}, {Leroy}, {Walter}, {Brinks}, {de
  Blok}, {Madore}  \& {Thornley}}{{Bigiel} et~al.}{2008}]{BigielEtAl2008}
{Bigiel} F.,  {Leroy} A.,  {Walter} F.,  {Brinks} E.,  {de Blok} W.~J.~G.,
  {Madore} B.,   {Thornley} M.~D.,  2008, \mn@doi [\aj]
  {10.1088/0004-6256/136/6/2846}, \href
  {http://adsabs.harvard.edu/abs/2008AJ....136.2846B} {136, 2846}

\bibitem[\protect\citeauthoryear{{Bohlin}, {Savage}  \& {Drake}}{{Bohlin}
  et~al.}{1978}]{BohlinSavageDrake1978}
{Bohlin} R.~C.,  {Savage} B.~D.,   {Drake} J.~F.,  1978, \mn@doi [\apj]
  {10.1086/156357}, \href
  {https://ui.adsabs.harvard.edu/abs/1978ApJ...224..132B} {224, 132}

\bibitem[\protect\citeauthoryear{Bouchut, Klingenberg  \& Waagan}{Bouchut
  et~al.}{2007}]{Bouchut2007}
Bouchut F.,  Klingenberg C.,   Waagan K.,  2007, \mn@doi [Numer. Math.]
  {10.1007/s00211-007-0108-8}, 108, 7

\bibitem[\protect\citeauthoryear{Bouchut, Klingenberg  \& Waagan}{Bouchut
  et~al.}{2010}]{Bouchut2010}
Bouchut F.,  Klingenberg C.,   Waagan K.,  2010, \mn@doi [Numer. Math.]
  {http://dx.doi.org/10.1007/s00211-010-0289-4}, 115, 647

\bibitem[\protect\citeauthoryear{{Br{\"u}ggen} \& {Scannapieco}}{{Br{\"u}ggen}
  \& {Scannapieco}}{2016}]{BrueggenScannapieco2016}
{Br{\"u}ggen} M.,  {Scannapieco} E.,  2016, \mn@doi [\apj]
  {10.3847/0004-637X/822/1/31}, \href
  {https://ui.adsabs.harvard.edu/abs/2016ApJ...822...31B} {822, 31}

\bibitem[\protect\citeauthoryear{Burgers}{Burgers}{1948}]{Burgers1948}
Burgers J.,  1948, Elsevier, pp 171 -- 199,
  \mn@doi{https://doi.org/10.1016/S0065-2156(08)70100-5}, \url
  {http://www.sciencedirect.com/science/article/pii/S0065215608701005}

\bibitem[\protect\citeauthoryear{{Chandrasekhar}}{{Chandrasekhar}}{1961}]{Chandrasekhar1961}
{Chandrasekhar} S.,  1961, {Hydrodynamic and hydromagnetic stability}.
Clarendon Press

\bibitem[\protect\citeauthoryear{{Clark}, {Glover}  \& {Klessen}}{{Clark}
  et~al.}{2012}]{ClarkGloverKlessen2012}
{Clark} P.~C.,  {Glover} S.~C.~O.,   {Klessen} R.~S.,  2012, \mn@doi [\mnras]
  {10.1111/j.1365-2966.2011.20087.x}, \href
  {http://adsabs.harvard.edu/abs/2012MNRAS.420..745C} {420, 745}

\bibitem[\protect\citeauthoryear{{Cooper}, {Bicknell}, {Sutherland }  \&
  {Bland-Hawthorn}}{{Cooper} et~al.}{2009}]{CooperEtAl2009}
{Cooper} J.~L.,  {Bicknell} G.~V.,  {Sutherland } R.~S.,   {Bland-Hawthorn} J.,
   2009, \mn@doi [\apj] {10.1088/0004-637X/703/1/330}, \href
  {https://ui.adsabs.harvard.edu/abs/2009ApJ...703..330C} {703, 330}

\bibitem[\protect\citeauthoryear{{Cottle}, {Scannapieco}, {Br{\"u}ggen},
  {Banda-Barrag{\'a}n}  \& {Federrath}}{{Cottle} et~al.}{2020}]{CottleEtAl2020}
{Cottle} J.,  {Scannapieco} E.,  {Br{\"u}ggen} M.,  {Banda-Barrag{\'a}n} W.,
  {Federrath} C.,  2020, \mn@doi [\apj] {10.3847/1538-4357/ab76d1}, \href
  {https://ui.adsabs.harvard.edu/abs/2020ApJ...892...59C} {892, 59}

\bibitem[\protect\citeauthoryear{{Crutcher}}{{Crutcher}}{2012}]{Crutcher2012}
{Crutcher} R.~M.,  2012, \mn@doi [\araa] {10.1146/annurev-astro-081811-125514},
  \href {http://adsabs.harvard.edu/abs/2012ARA%26A..50...29C} {50, 29}

\bibitem[\protect\citeauthoryear{{Dalgarno} \& {McCray}}{{Dalgarno} \&
  {McCray}}{1972}]{DalgarnoMcCray1972}
{Dalgarno} A.,  {McCray} R.~A.,  1972, \mn@doi [\araa]
  {10.1146/annurev.aa.10.090172.002111}, \href
  {https://ui.adsabs.harvard.edu/abs/1972ARA&A..10..375D} {10, 375}

\bibitem[\protect\citeauthoryear{{Dashyan} \& {Dubois}}{{Dashyan} \&
  {Dubois}}{2020}]{DashanDubois2020}
{Dashyan} G.,  {Dubois} Y.,  2020, \mn@doi [\aap]
  {10.1051/0004-6361/201936339}, \href
  {https://ui.adsabs.harvard.edu/abs/2020A&A...638A.123D} {638, A123}

\bibitem[\protect\citeauthoryear{{Di Teodoro}, {McClure-Griffiths}, {Lockman},
  {Denbo}, {Endsley}, {Ford}  \& {Harrington}}{{Di Teodoro}
  et~al.}{2018}]{DiTeodoroEtAl2018}
{Di Teodoro} E.~M.,  {McClure-Griffiths} N.~M.,  {Lockman} F.~J.,  {Denbo}
  S.~R.,  {Endsley} R.,  {Ford} H.~A.,   {Harrington} K.,  2018, \mn@doi [\apj]
  {10.3847/1538-4357/aaad6a}, \href
  {https://ui.adsabs.harvard.edu/abs/2018ApJ...855...33D} {855, 33}

\bibitem[\protect\citeauthoryear{{Di Teodoro} et~al.,}{{Di Teodoro}
  et~al.}{2019}]{DiTeodoroEtAl2019}
{Di Teodoro} E.~M.,  et~al., 2019, \mn@doi [\apjl] {10.3847/2041-8213/ab4fe9},
  \href {https://ui.adsabs.harvard.edu/abs/2019ApJ...885L..32D} {885, L32}

\bibitem[\protect\citeauthoryear{{Di Teodoro}, {McClure-Griffiths}, {Lockman}
  \& {Armillotta}}{{Di Teodoro} et~al.}{2020}]{DiTeodoroEtAl2020}
{Di Teodoro} E.~M.,  {McClure-Griffiths} N.~M.,  {Lockman} F.~J.,
  {Armillotta} L.,  2020, \mn@doi [\nat] {10.1038/s41586-020-2595-z}, \href
  {https://ui.adsabs.harvard.edu/abs/2020Natur.584..364D} {584, 364}

\bibitem[\protect\citeauthoryear{{Draine}}{{Draine}}{1978}]{Draine1978}
{Draine} B.~T.,  1978, \mn@doi [\apjs] {10.1086/190513}, \href
  {http://adsabs.harvard.edu/abs/1978ApJS...36..595D} {36, 595}

\bibitem[\protect\citeauthoryear{{Draine}}{{Draine}}{2011}]{Draine2011}
{Draine} B.~T.,  2011, {Physics of the Interstellar and Intergalactic Medium}

\bibitem[\protect\citeauthoryear{Dubey et~al.,}{Dubey
  et~al.}{2008}]{DubeyEtAl2008}
Dubey A.,  et~al., 2008, in Numerical Modeling of Space Plasma Flows: Astronum
  2007. p.~145

\bibitem[\protect\citeauthoryear{{Dursi} \& {Pfrommer}}{{Dursi} \&
  {Pfrommer}}{2008}]{DursiPfrommer2008}
{Dursi} L.~J.,  {Pfrommer} C.,  2008, \mn@doi [\apj] {10.1086/529371}, \href
  {https://ui.adsabs.harvard.edu/abs/2008ApJ...677..993D} {677, 993}

\bibitem[\protect\citeauthoryear{{Emerick}, {Bryan}  \& {Mac Low}}{{Emerick}
  et~al.}{2019}]{EmerickBryanMacLow2019}
{Emerick} A.,  {Bryan} G.~L.,   {Mac Low} M.-M.,  2019, \mn@doi [\mnras]
  {10.1093/mnras/sty2689}, \href
  {https://ui.adsabs.harvard.edu/abs/2019MNRAS.482.1304E} {482, 1304}

\bibitem[\protect\citeauthoryear{{Fielding}, {Quataert}  \&
  {Martizzi}}{{Fielding} et~al.}{2018}]{FieldingQuataertMartizzi2018}
{Fielding} D.,  {Quataert} E.,   {Martizzi} D.,  2018, \mn@doi [\mnras]
  {10.1093/mnras/sty2466}, \href
  {https://ui.adsabs.harvard.edu/abs/2018MNRAS.481.3325F} {481, 3325}

\bibitem[\protect\citeauthoryear{{F{\"o}rster Schreiber} \&
  {Wuyts}}{{F{\"o}rster Schreiber} \&
  {Wuyts}}{2020}]{FoersterSchreiberStijn2020}
{F{\"o}rster Schreiber} N.~M.,  {Wuyts} S.,  2020, \mn@doi [\araa]
  {10.1146/annurev-astro-032620-021910}, \href
  {https://ui.adsabs.harvard.edu/abs/2020ARA&A..58..661F} {58, 661}

\bibitem[\protect\citeauthoryear{{F{\"o}rster Schreiber} et~al.,}{{F{\"o}rster
  Schreiber} et~al.}{2019}]{FoersterSchreiberEtAl2019}
{F{\"o}rster Schreiber} N.~M.,  et~al., 2019, \mn@doi [\apj]
  {10.3847/1538-4357/ab0ca2}, \href
  {https://ui.adsabs.harvard.edu/abs/2019ApJ...875...21F} {875, 21}

\bibitem[\protect\citeauthoryear{{Fryxell} et~al.,}{{Fryxell}
  et~al.}{2000}]{FLASH00}
{Fryxell} B.,  et~al., 2000, \apjs, \href
  {http://cdsads.u-strasbg.fr/cgi-bin/nph-bib_query?bibcode=2000ApJS..131..273F&amp;db_key=AST}
  {131, 273}

\bibitem[\protect\citeauthoryear{{Gatto} et~al.,}{{Gatto}
  et~al.}{2017}]{GattoEtAl2017}
{Gatto} A.,  et~al., 2017, \mn@doi [\mnras] {10.1093/mnras/stw3209}, \href
  {http://adsabs.harvard.edu/abs/2017MNRAS.466.1903G} {466, 1903}

\bibitem[\protect\citeauthoryear{{Girichidis}, {Konstandin}, {Whitworth}  \&
  {Klessen}}{{Girichidis} et~al.}{2014}]{GirichidisEtAl2014}
{Girichidis} P.,  {Konstandin} L.,  {Whitworth} A.~P.,   {Klessen} R.~S.,
  2014, \mn@doi [\apj] {10.1088/0004-637X/781/2/91}, \href
  {http://adsabs.harvard.edu/abs/2014ApJ...781...91G} {781, 91}

\bibitem[\protect\citeauthoryear{{Girichidis} et~al.,}{{Girichidis}
  et~al.}{2016a}]{GirichidisEtAl2016b}
{Girichidis} P.,  et~al., 2016a, \mn@doi [\mnras] {10.1093/mnras/stv2742},
  \href {http://adsabs.harvard.edu/abs/2016MNRAS.456.3432G} {456, 3432}

\bibitem[\protect\citeauthoryear{{Girichidis} et~al.,}{{Girichidis}
  et~al.}{2016b}]{GirichidisEtAl2016a}
{Girichidis} P.,  et~al., 2016b, \mn@doi [\apjl] {10.3847/2041-8205/816/2/L19},
  \href {http://adsabs.harvard.edu/abs/2016ApJ...816L..19G} {816, L19}

\bibitem[\protect\citeauthoryear{{Girichidis}, {Naab}, {Hanasz}  \&
  {Walch}}{{Girichidis} et~al.}{2018}]{GirichidisEtAl2018a}
{Girichidis} P.,  {Naab} T.,  {Hanasz} M.,   {Walch} S.,  2018, \mn@doi
  [\mnras] {10.1093/mnras/sty1653}, \href
  {http://adsabs.harvard.edu/abs/2018MNRAS.479.3042G} {479, 3042}

\bibitem[\protect\citeauthoryear{{Girichidis} et~al.,}{{Girichidis}
  et~al.}{2020}]{GirichidisEtAl2020b}
{Girichidis} P.,  et~al., 2020, \mn@doi [\ssr] {10.1007/s11214-020-00693-8},
  \href {https://ui.adsabs.harvard.edu/abs/2020SSRv..216...68G} {216, 68}

\bibitem[\protect\citeauthoryear{{Glover} \& {Clark}}{{Glover} \&
  {Clark}}{2012a}]{GloverClark2012a}
{Glover} S.~C.~O.,  {Clark} P.~C.,  2012a, \mn@doi [\mnras]
  {10.1111/j.1365-2966.2011.19648.x}, \href
  {http://adsabs.harvard.edu/abs/2012MNRAS.421....9G} {421, 9}

\bibitem[\protect\citeauthoryear{{Glover} \& {Clark}}{{Glover} \&
  {Clark}}{2012b}]{GloverClark2012b}
{Glover} S.~C.~O.,  {Clark} P.~C.,  2012b, \mn@doi [\mnras]
  {10.1111/j.1365-2966.2011.20260.x}, \href
  {http://adsabs.harvard.edu/abs/2012MNRAS.421..116G} {421, 116}

\bibitem[\protect\citeauthoryear{{Glover} \& {Mac Low}}{{Glover} \& {Mac
  Low}}{2007}]{GloverMacLow2007a}
{Glover} S.~C.~O.,  {Mac Low} M.-M.,  2007, \mn@doi [\apjs] {10.1086/512238},
  \href {http://cdsads.u-strasbg.fr/abs/2007ApJS..169..239G} {169, 239}

\bibitem[\protect\citeauthoryear{{Glover}, {Federrath}, {Mac Low}  \&
  {Klessen}}{{Glover} et~al.}{2010}]{GloverEtAl2010}
{Glover} S.~C.~O.,  {Federrath} C.,  {Mac Low} M.-M.,   {Klessen} R.~S.,  2010,
  \mn@doi [\mnras] {10.1111/j.1365-2966.2009.15718.x}, \href
  {http://adsabs.harvard.edu/abs/2010MNRAS.404....2G} {404, 2}

\bibitem[\protect\citeauthoryear{{Gnat} \& {Ferland}}{{Gnat} \&
  {Ferland}}{2012}]{GnatFerland2012}
{Gnat} O.,  {Ferland} G.~J.,  2012, \mn@doi [\apjs]
  {10.1088/0067-0049/199/1/20}, \href
  {http://adsabs.harvard.edu/abs/2012ApJS..199...20G} {199, 20}

\bibitem[\protect\citeauthoryear{{Goldsmith} \& {Langer}}{{Goldsmith} \&
  {Langer}}{1978}]{GoldsmithLanger1978}
{Goldsmith} P.~F.,  {Langer} W.~D.,  1978, \apj, \href
  {http://cdsads.u-strasbg.fr/cgi-bin/nph-bib_query?bibcode=1978ApJ...222..881G&db_key=AST}
  {222, 881}

\bibitem[\protect\citeauthoryear{{Goldsmith} \& {Pittard}}{{Goldsmith} \&
  {Pittard}}{2017}]{GoldsmithPittard2017}
{Goldsmith} K.~J.~A.,  {Pittard} J.~M.,  2017, \mn@doi [\mnras]
  {10.1093/mnras/stx1431}, \href
  {https://ui.adsabs.harvard.edu/abs/2017MNRAS.470.2427G} {470, 2427}

\bibitem[\protect\citeauthoryear{{Goldsmith} \& {Pittard}}{{Goldsmith} \&
  {Pittard}}{2018}]{GoldsmithPittard2018}
{Goldsmith} K.~J.~A.,  {Pittard} J.~M.,  2018, \mn@doi [\mnras]
  {10.1093/mnras/sty401}, \href
  {https://ui.adsabs.harvard.edu/abs/2018MNRAS.476.2209G} {476, 2209}

\bibitem[\protect\citeauthoryear{{Gronke} \& {Oh}}{{Gronke} \&
  {Oh}}{2018}]{GronkeOh2018}
{Gronke} M.,  {Oh} S.~P.,  2018, \mn@doi [\mnras] {10.1093/mnrasl/sly131},
  \href {https://ui.adsabs.harvard.edu/abs/2018MNRAS.480L.111G} {480, L111}

\bibitem[\protect\citeauthoryear{{Gronke} \& {Oh}}{{Gronke} \&
  {Oh}}{2020}]{GronkeOh2020a}
{Gronke} M.,  {Oh} S.~P.,  2020, \mn@doi [\mnras] {10.1093/mnras/stz3332},
  \href {https://ui.adsabs.harvard.edu/abs/2020MNRAS.492.1970G} {492, 1970}

\bibitem[\protect\citeauthoryear{{Gr{\o}nnow}, {Tepper-Garc{\'\i}a}  \& {Bland
  -Hawthorn}}{{Gr{\o}nnow} et~al.}{2018}]{GronnowEtAl2018}
{Gr{\o}nnow} A.,  {Tepper-Garc{\'\i}a} T.,   {Bland -Hawthorn} J.,  2018,
  \mn@doi [\apj] {10.3847/1538-4357/aada0e}, \href
  {https://ui.adsabs.harvard.edu/abs/2018ApJ...865...64G} {865, 64}

\bibitem[\protect\citeauthoryear{{Gutcke}, {Pakmor}, {Naab}  \&
  {Springel}}{{Gutcke} et~al.}{2020}]{GutckeEtAl2020}
{Gutcke} T.~A.,  {Pakmor} R.,  {Naab} T.,   {Springel} V.,  2020, arXiv
  e-prints, \href {https://ui.adsabs.harvard.edu/abs/2020arXiv201007311G} {p.
  arXiv:2010.07311}

\bibitem[\protect\citeauthoryear{{Habing}}{{Habing}}{1968}]{Habing1968}
{Habing} H.~J.,  1968, \bain, \href
  {http://adsabs.harvard.edu/abs/1968BAN....19..421H} {19, 421}

\bibitem[\protect\citeauthoryear{{Han}}{{Han}}{2017}]{Han2017}
{Han} J.~L.,  2017, \mn@doi [\araa] {10.1146/annurev-astro-091916-055221},
  \href {https://ui.adsabs.harvard.edu/abs/2017ARA&A..55..111H} {55, 111}

\bibitem[\protect\citeauthoryear{{Haverkorn}}{{Haverkorn}}{2015}]{Haverkorn2015}
{Haverkorn} M.,  2015, in {Lazarian} A.,  {de Gouveia Dal Pino} E.~M.,
  {Melioli} C.,  eds,  Astrophysics and Space Science Library Vol. 407,
  Magnetic Fields in Diffuse Media. p.~483 (\mn@eprint {arXiv} {1406.0283}),
  \mn@doi{10.1007/978-3-662-44625-6\_17}

\bibitem[\protect\citeauthoryear{{Hollenbach} \& {McKee}}{{Hollenbach} \&
  {McKee}}{1989}]{Hollenbach89}
{Hollenbach} D.,  {McKee} C.~F.,  1989, \mn@doi [\apj] {10.1086/167595}, \href
  {http://cdsads.u-strasbg.fr/cgi-bin/nph-bib_query?bibcode=1989ApJ...342..306H&db_key=AST}
  {342, 306}

\bibitem[\protect\citeauthoryear{{Hollenbach}, {Werner}  \&
  {Salpeter}}{{Hollenbach} et~al.}{1971}]{HollenbachWernerSalpeter1971}
{Hollenbach} D.~J.,  {Werner} M.~W.,   {Salpeter} E.~E.,  1971, \mn@doi [\apj]
  {10.1086/150755}, \href
  {https://ui.adsabs.harvard.edu/abs/1971ApJ...163..165H} {163, 165}

\bibitem[\protect\citeauthoryear{{Hu}}{{Hu}}{2019}]{Hu2019}
{Hu} C.-Y.,  2019, \mn@doi [\mnras] {10.1093/mnras/sty3252}, \href
  {https://ui.adsabs.harvard.edu/abs/2019MNRAS.483.3363H} {483, 3363}

\bibitem[\protect\citeauthoryear{{Hu}, {Naab}, {Walch}, {Glover}  \&
  {Clark}}{{Hu} et~al.}{2016}]{HuEtAl2016}
{Hu} C.-Y.,  {Naab} T.,  {Walch} S.,  {Glover} S.~C.~O.,   {Clark} P.~C.,
  2016, \mn@doi [\mnras] {10.1093/mnras/stw544}, \href
  {http://adsabs.harvard.edu/abs/2016MNRAS.458.3528H} {458, 3528}

\bibitem[\protect\citeauthoryear{{Jeans}}{{Jeans}}{1902}]{Jeans1902}
{Jeans} J.~H.,  1902, \mn@doi [Royal Society of London Philosophical
  Transactions Series A] {10.1098/rsta.1902.0012}, \href
  {http://adsabs.harvard.edu/abs/1902RSPTA.199....1J} {199, 1}

\bibitem[\protect\citeauthoryear{{Ji}, {Oh}  \& {Masterson}}{{Ji}
  et~al.}{2019}]{JiOhMasterson2019}
{Ji} S.,  {Oh} S.~P.,   {Masterson} P.,  2019, \mn@doi [\mnras]
  {10.1093/mnras/stz1248}, \href
  {https://ui.adsabs.harvard.edu/abs/2019MNRAS.487..737J} {487, 737}

\bibitem[\protect\citeauthoryear{{Johansson} \& {Ziegler}}{{Johansson} \&
  {Ziegler}}{2013}]{JohanssonZiegler2013}
{Johansson} E. P.~G.,  {Ziegler} U.,  2013, \mn@doi [\apj]
  {10.1088/0004-637X/766/1/45}, \href
  {https://ui.adsabs.harvard.edu/abs/2013ApJ...766...45J} {766, 45}

\bibitem[\protect\citeauthoryear{{Kim} \& {Ostriker}}{{Kim} \&
  {Ostriker}}{2018}]{KimOstriker2018}
{Kim} C.-G.,  {Ostriker} E.~C.,  2018, \mn@doi [\apj]
  {10.3847/1538-4357/aaa5ff}, \href
  {http://adsabs.harvard.edu/abs/2018ApJ...853..173K} {853, 173}

\bibitem[\protect\citeauthoryear{{Klein}, {McKee}  \& {Colella}}{{Klein}
  et~al.}{1994}]{KleinMcKeeColella1994}
{Klein} R.~I.,  {McKee} C.~F.,   {Colella} P.,  1994, \mn@doi [\apj]
  {10.1086/173554}, \href
  {https://ui.adsabs.harvard.edu/abs/1994ApJ...420..213K} {420, 213}

\bibitem[\protect\citeauthoryear{{Klessen} \& {Glover}}{{Klessen} \&
  {Glover}}{2016}]{KlessenGlover2016}
{Klessen} R.~S.,  {Glover} S.~C.~O.,  2016, \mn@doi [Star Formation in Galaxy
  Evolution: Connecting Numerical Models to Reality, Saas-Fee Advanced Course,
  Volume 43.~ISBN 978-3-662-47889-9.~Springer-Verlag Berlin Heidelberg, 2016,
  p.~85] {10.1007/978-3-662-47890-5_2}, \href
  {http://adsabs.harvard.edu/abs/2016SAAS...43...85K} {43, 85}

\bibitem[\protect\citeauthoryear{{Krieger} et~al.,}{{Krieger}
  et~al.}{2019}]{KriegerEtAl2019}
{Krieger} N.,  et~al., 2019, \mn@doi [\apj] {10.3847/1538-4357/ab2d9c}, \href
  {https://ui.adsabs.harvard.edu/abs/2019ApJ...881...43K} {881, 43}

\bibitem[\protect\citeauthoryear{Lautrup}{Lautrup}{2011}]{Lautrup2011}
Lautrup B.,  2011, Physics of continuous matter: exotic and everyday phenomena
  in the macroscopic world.
CRC press

\bibitem[\protect\citeauthoryear{{Leroy} et~al.,}{{Leroy}
  et~al.}{2015}]{LeroyEtAl2015}
{Leroy} A.~K.,  et~al., 2015, \mn@doi [\apj] {10.1088/0004-637X/814/2/83},
  \href {https://ui.adsabs.harvard.edu/abs/2015ApJ...814...83L} {814, 83}

\bibitem[\protect\citeauthoryear{{Li}, {Bryan}  \& {Ostriker}}{{Li}
  et~al.}{2017}]{LiBryanOstriker2017}
{Li} M.,  {Bryan} G.~L.,   {Ostriker} J.~P.,  2017, \mn@doi [\apj]
  {10.3847/1538-4357/aa7263}, \href
  {http://adsabs.harvard.edu/abs/2017ApJ...841..101L} {841, 101}

\bibitem[\protect\citeauthoryear{{Lockman}, {Di Teodoro}  \&
  {McClure-Griffiths}}{{Lockman}
  et~al.}{2020}]{LockmanDiTeodoroMcClureGriffiths2020}
{Lockman} F.~J.,  {Di Teodoro} E.~M.,   {McClure-Griffiths} N.~M.,  2020,
  \mn@doi [\apj] {10.3847/1538-4357/ab55d8}, \href
  {https://ui.adsabs.harvard.edu/abs/2020ApJ...888...51L} {888, 51}

\bibitem[\protect\citeauthoryear{{Lutz} et~al.,}{{Lutz}
  et~al.}{2020}]{LutzEtAl2020}
{Lutz} D.,  et~al., 2020, \mn@doi [\aap] {10.1051/0004-6361/201936803}, \href
  {https://ui.adsabs.harvard.edu/abs/2020A&A...633A.134L} {633, A134}

\bibitem[\protect\citeauthoryear{{Marinacci}, {Binney}, {Fraternali}, {Nipoti},
  {Ciotti}  \& {Londrillo}}{{Marinacci} et~al.}{2010}]{MarinacciEtAl2010}
{Marinacci} F.,  {Binney} J.,  {Fraternali} F.,  {Nipoti} C.,  {Ciotti} L.,
  {Londrillo} P.,  2010, \mn@doi [\mnras] {10.1111/j.1365-2966.2010.16352.x},
  \href {https://ui.adsabs.harvard.edu/abs/2010MNRAS.404.1464M} {404, 1464}

\bibitem[\protect\citeauthoryear{{Martizzi}, {Fielding}, {Faucher-Gigu{\`e}re}
  \& {Quataert}}{{Martizzi} et~al.}{2016}]{MartizziEtAl2016}
{Martizzi} D.,  {Fielding} D.,  {Faucher-Gigu{\`e}re} C.-A.,   {Quataert} E.,
  2016, \mn@doi [\mnras] {10.1093/mnras/stw745}, \href
  {http://adsabs.harvard.edu/abs/2016MNRAS.459.2311M} {459, 2311}

\bibitem[\protect\citeauthoryear{{Mathis}, {Mezger}  \& {Panagia}}{{Mathis}
  et~al.}{1983}]{MathisMezgerPanagia1983}
{Mathis} J.~S.,  {Mezger} P.~G.,   {Panagia} N.,  1983, \aap, \href
  {http://adsabs.harvard.edu/abs/1983A%26A...128..212M} {128, 212}

\bibitem[\protect\citeauthoryear{{McClure-Griffiths}, {Green}, {Hill},
  {Lockman}, {Dickey}, {Gaensler}  \& {Green}}{{McClure-Griffiths}
  et~al.}{2013}]{McClureGriffiths2013}
{McClure-Griffiths} N.~M.,  {Green} J.~A.,  {Hill} A.~S.,  {Lockman} F.~J.,
  {Dickey} J.~M.,  {Gaensler} B.~M.,   {Green} A.~J.,  2013, \mn@doi [\apjl]
  {10.1088/2041-8205/770/1/L4}, \href
  {https://ui.adsabs.harvard.edu/abs/2013ApJ...770L...4M} {770, L4}

\bibitem[\protect\citeauthoryear{{McCourt}, {O'Leary}, {Madigan}  \&
  {Quataert}}{{McCourt} et~al.}{2015}]{McCourtEtAl2015}
{McCourt} M.,  {O'Leary} R.~M.,  {Madigan} A.-M.,   {Quataert} E.,  2015,
  \mn@doi [\mnras] {10.1093/mnras/stv355}, \href
  {https://ui.adsabs.harvard.edu/abs/2015MNRAS.449....2M} {449, 2}

\bibitem[\protect\citeauthoryear{{McCourt}, {Oh}, {O'Leary}  \&
  {Madigan}}{{McCourt} et~al.}{2018}]{McCourtEtAl2018}
{McCourt} M.,  {Oh} S.~P.,  {O'Leary} R.,   {Madigan} A.-M.,  2018, \mn@doi
  [\mnras] {10.1093/mnras/stx2687}, \href
  {https://ui.adsabs.harvard.edu/abs/2018MNRAS.473.5407M} {473, 5407}

\bibitem[\protect\citeauthoryear{{McKee} \& {Ostriker}}{{McKee} \&
  {Ostriker}}{1977}]{McKeeOstriker1977}
{McKee} C.~F.,  {Ostriker} J.~P.,  1977, \mn@doi [\apj] {10.1086/155667}, \href
  {http://adsabs.harvard.edu/abs/1977ApJ...218..148M} {218, 148}

\bibitem[\protect\citeauthoryear{{Mellema}, {Kurk}  \&
  {R{\"o}ttgering}}{{Mellema} et~al.}{2002}]{MellemaKurkRoettgering2002}
{Mellema} G.,  {Kurk} J.~D.,   {R{\"o}ttgering} H.~J.~A.,  2002, \mn@doi [\aap]
  {10.1051/0004-6361:20021408}, \href
  {https://ui.adsabs.harvard.edu/abs/2002A&A...395L..13M} {395, L13}

\bibitem[\protect\citeauthoryear{Meshkov}{Meshkov}{1969}]{Meshkov1969}
Meshkov E.~E.,  1969, \mn@doi [Fluid Dynamics] {10.1007/BF01015969}, 4, 101

\bibitem[\protect\citeauthoryear{{Micic}, {Glover}, {Federrath}  \&
  {Klessen}}{{Micic} et~al.}{2012}]{MicicEtAl2012}
{Micic} M.,  {Glover} S.~C.~O.,  {Federrath} C.,   {Klessen} R.~S.,  2012,
  \mn@doi [\mnras] {10.1111/j.1365-2966.2012.20477.x}, \href
  {http://adsabs.harvard.edu/abs/2012MNRAS.421.2531M} {421, 2531}

\bibitem[\protect\citeauthoryear{{Murray}, {White}, {Blondin}  \&
  {Lin}}{{Murray} et~al.}{1993}]{MurrayEtAl1993}
{Murray} S.~D.,  {White} S. D.~M.,  {Blondin} J.~M.,   {Lin} D. N.~C.,  1993,
  \mn@doi [\apj] {10.1086/172540}, \href
  {https://ui.adsabs.harvard.edu/abs/1993ApJ...407..588M} {407, 588}

\bibitem[\protect\citeauthoryear{{Naab} \& {Ostriker}}{{Naab} \&
  {Ostriker}}{2017}]{NaabOstriker2017}
{Naab} T.,  {Ostriker} J.~P.,  2017, \mn@doi [\araa]
  {10.1146/annurev-astro-081913-040019}, \href
  {http://adsabs.harvard.edu/abs/2017ARA%26A..55...59N} {55, 59}

\bibitem[\protect\citeauthoryear{{Nakamura}, {McKee}, {Klein}  \&
  {Fisher}}{{Nakamura} et~al.}{2006}]{NakamuraEtAl2006}
{Nakamura} F.,  {McKee} C.~F.,  {Klein} R.~I.,   {Fisher} R.~T.,  2006, \mn@doi
  [\apjs] {10.1086/501530}, \href
  {https://ui.adsabs.harvard.edu/abs/2006ApJS..164..477N} {164, 477}

\bibitem[\protect\citeauthoryear{{Nelson} \& {Langer}}{{Nelson} \&
  {Langer}}{1997}]{NelsonLanger1997}
{Nelson} R.~P.,  {Langer} W.~D.,  1997, \apj, \href
  {http://adsabs.harvard.edu/abs/1997ApJ...482..796N} {482, 796}

\bibitem[\protect\citeauthoryear{{Olson}, {MacNeice}, {Fryxell}, {Ricker},
  {Timmes}  \& {Zingale}}{{Olson} et~al.}{1999}]{PARAMESH99}
{Olson} K.~M.,  {MacNeice} P.,  {Fryxell} B.,  {Ricker} P.,  {Timmes} F.~X.,
  {Zingale} M.,  1999, Bulletin of the American Astronomical Society, \href
  {http://cdsads.u-strasbg.fr/cgi-bin/nph-bib_query?bibcode=1999AAS...195.4203O&amp;db_key=AST}
  {31, 1430}

\bibitem[\protect\citeauthoryear{{Ossenkopf} \& {Henning}}{{Ossenkopf} \&
  {Henning}}{1994}]{OssenkopfHenning1994}
{Ossenkopf} V.,  {Henning} T.,  1994, \aap, \href
  {http://adsabs.harvard.edu/abs/1994A%26A...291..943O} {291, 943}

\bibitem[\protect\citeauthoryear{{Peters} et~al.,}{{Peters}
  et~al.}{2017}]{PetersEtAl2017a}
{Peters} T.,  et~al., 2017, \mn@doi [\mnras] {10.1093/mnras/stw3216}, \href
  {http://adsabs.harvard.edu/abs/2017MNRAS.466.3293P} {466, 3293}

\bibitem[\protect\citeauthoryear{{Pittard} \& {Parkin}}{{Pittard} \&
  {Parkin}}{2016}]{PittardParkin2016}
{Pittard} J.~M.,  {Parkin} E.~R.,  2016, \mn@doi [\mnras]
  {10.1093/mnras/stw025}, \href
  {https://ui.adsabs.harvard.edu/abs/2016MNRAS.457.4470P} {457, 4470}

\bibitem[\protect\citeauthoryear{{Pittard}, {Dyson}, {Falle}  \&
  {Hartquist}}{{Pittard} et~al.}{2005}]{PittardEtAl2005}
{Pittard} J.~M.,  {Dyson} J.~E.,  {Falle} S.~A.~E.~G.,   {Hartquist} T.~W.,
  2005, \mn@doi [\mnras] {10.1111/j.1365-2966.2005.09268.x}, \href
  {https://ui.adsabs.harvard.edu/abs/2005MNRAS.361.1077P} {361, 1077}

\bibitem[\protect\citeauthoryear{{Rathjen} et~al.,}{{Rathjen}
  et~al.}{2021}]{RathjenEtAl2021}
{Rathjen} T.-E.,  et~al., 2021, \mn@doi [\mnras] {10.1093/mnras/stab900}, \href
  {https://ui.adsabs.harvard.edu/abs/2021MNRAS.504.1039R} {504, 1039}

\bibitem[\protect\citeauthoryear{Richtmyer}{Richtmyer}{1960}]{Richtmyer1960}
Richtmyer R.~D.,  1960, \mn@doi [Communications on Pure and Applied
  Mathematics] {https://doi.org/10.1002/cpa.3160130207}, 13, 297

\bibitem[\protect\citeauthoryear{{Roy}, {Nath}, {Sharma}  \&
  {Shchekinov}}{{Roy} et~al.}{2016}]{RoyEtAl2016}
{Roy} A.,  {Nath} B.~B.,  {Sharma} P.,   {Shchekinov} Y.,  2016, \mn@doi
  [\mnras] {10.1093/mnras/stw2127}, \href
  {https://ui.adsabs.harvard.edu/abs/2016MNRAS.463.2296R} {463, 2296}

\bibitem[\protect\citeauthoryear{{Rupke}}{{Rupke}}{2018}]{Rupke2018}
{Rupke} D.,  2018, \mn@doi [Galaxies] {10.3390/galaxies6040138}, \href
  {https://ui.adsabs.harvard.edu/abs/2018Galax...6..138R} {6, 138}

\bibitem[\protect\citeauthoryear{{Salak}, {Nakai}, {Sorai}  \&
  {Miyamoto}}{{Salak} et~al.}{2020}]{SalakEtAl2020}
{Salak} D.,  {Nakai} N.,  {Sorai} K.,   {Miyamoto} Y.,  2020, \mn@doi [\apj]
  {10.3847/1538-4357/abb134}, \href
  {https://ui.adsabs.harvard.edu/abs/2020ApJ...901..151S} {901, 151}

\bibitem[\protect\citeauthoryear{{Scannapieco} \& {Br{\"u}ggen}}{{Scannapieco}
  \& {Br{\"u}ggen}}{2015}]{ScannapiecoBrueggen2015}
{Scannapieco} E.,  {Br{\"u}ggen} M.,  2015, \mn@doi [\apj]
  {10.1088/0004-637X/805/2/158}, \href
  {https://ui.adsabs.harvard.edu/abs/2015ApJ...805..158S} {805, 158}

\bibitem[\protect\citeauthoryear{{Schneider} \& {Robertson}}{{Schneider} \&
  {Robertson}}{2015}]{SchneiderRobertson2015}
{Schneider} E.~E.,  {Robertson} B.~E.,  2015, \mn@doi [\apjs]
  {10.1088/0067-0049/217/2/24}, \href
  {https://ui.adsabs.harvard.edu/abs/2015ApJS..217...24S} {217, 24}

\bibitem[\protect\citeauthoryear{{Schneider} \& {Robertson}}{{Schneider} \&
  {Robertson}}{2017}]{SchneiderRobertson2017}
{Schneider} E.~E.,  {Robertson} B.~E.,  2017, \mn@doi [\apj]
  {10.3847/1538-4357/834/2/144}, \href
  {https://ui.adsabs.harvard.edu/abs/2017ApJ...834..144S} {834, 144}

\bibitem[\protect\citeauthoryear{{Schneider}, {Ostriker}, {Robertson}  \&
  {Thompson}}{{Schneider} et~al.}{2020}]{SchneiderEtAl2020}
{Schneider} E.~E.,  {Ostriker} E.~C.,  {Robertson} B.~E.,   {Thompson} T.~A.,
  2020, \mn@doi [\apj] {10.3847/1538-4357/ab8ae8}, \href
  {https://ui.adsabs.harvard.edu/abs/2020ApJ...895...43S} {895, 43}

\bibitem[\protect\citeauthoryear{{Seifried} et~al.,}{{Seifried}
  et~al.}{2017}]{SeifriedEtAl2017}
{Seifried} D.,  et~al., 2017, \mn@doi [\mnras] {10.1093/mnras/stx2343}, \href
  {http://adsabs.harvard.edu/abs/2017MNRAS.472.4797S} {472, 4797}

\bibitem[\protect\citeauthoryear{{Seifried}, {Haid}, {Walch}, {Borchert}  \&
  {Bisbas}}{{Seifried} et~al.}{2020}]{SeifriedEtAl2020Zoom}
{Seifried} D.,  {Haid} S.,  {Walch} S.,  {Borchert} E.~M.~A.,   {Bisbas} T.~G.,
   2020, \mn@doi [\mnras] {10.1093/mnras/stz3563}, \href
  {https://ui.adsabs.harvard.edu/abs/2020MNRAS.492.1465S} {492, 1465}

\bibitem[\protect\citeauthoryear{{Sembach}, {Howk}, {Ryans}  \&
  {Keenan}}{{Sembach} et~al.}{2000}]{SembachEtAl2000}
{Sembach} K.~R.,  {Howk} J.~C.,  {Ryans} R.~S.~I.,   {Keenan} F.~P.,  2000,
  \mn@doi [\apj] {10.1086/308173}, \href
  {http://adsabs.harvard.edu/abs/2000ApJ...528..310S} {528, 310}

\bibitem[\protect\citeauthoryear{{Semenov}, {Kravtsov}  \&
  {Caprioli}}{{Semenov} et~al.}{2020}]{SemenovKravtsovCaprioli2020}
{Semenov} V.~A.,  {Kravtsov} A.~V.,   {Caprioli} D.,  2020, arXiv e-prints,
  \href {https://ui.adsabs.harvard.edu/abs/2020arXiv201201427S} {p.
  arXiv:2012.01427}

\bibitem[\protect\citeauthoryear{{Shu}}{{Shu}}{1992}]{ShuAstroGas1992}
{Shu} F.~H.,  1992, {The physics of astrophysics. Volume II: Gas dynamics.}

\bibitem[\protect\citeauthoryear{{Simpson}, {Pakmor}, {Marinacci}, {Pfrommer},
  {Springel}, {Glover}, {Clark}  \& {Smith}}{{Simpson}
  et~al.}{2016}]{SimpsonEtAl2016}
{Simpson} C.~M.,  {Pakmor} R.,  {Marinacci} F.,  {Pfrommer} C.,  {Springel} V.,
   {Glover} S.~C.~O.,  {Clark} P.~C.,   {Smith} R.~J.,  2016, \mn@doi [\apjl]
  {10.3847/2041-8205/827/2/L29}, \href
  {http://adsabs.harvard.edu/abs/2016ApJ...827L..29S} {827, L29}

\bibitem[\protect\citeauthoryear{{Smith}, {Bryan}, {Somerville}, {Hu},
  {Teyssier}, {Burkhart}  \& {Hernquist}}{{Smith}
  et~al.}{2020a}]{SmithMEtAl2020}
{Smith} M.~C.,  {Bryan} G.~L.,  {Somerville} R.~S.,  {Hu} C.-Y.,  {Teyssier}
  R.,  {Burkhart} B.,   {Hernquist} L.,  2020a, arXiv e-prints, \href
  {https://ui.adsabs.harvard.edu/abs/2020arXiv200911309S} {p. arXiv:2009.11309}

\bibitem[\protect\citeauthoryear{{Smith} et~al.,}{{Smith}
  et~al.}{2020b}]{SmithREtAl2020}
{Smith} R.~J.,  et~al., 2020b, \mn@doi [\mnras] {10.1093/mnras/stz3328}, \href
  {https://ui.adsabs.harvard.edu/abs/2020MNRAS.492.1594S} {492, 1594}

\bibitem[\protect\citeauthoryear{{Sparre}, {Pfrommer}  \&
  {Vogelsberger}}{{Sparre} et~al.}{2019}]{SparrePfrommerVogelsberger2019}
{Sparre} M.,  {Pfrommer} C.,   {Vogelsberger} M.,  2019, \mn@doi [\mnras]
  {10.1093/mnras/sty3063}, \href
  {https://ui.adsabs.harvard.edu/abs/2019MNRAS.482.5401S} {482, 5401}

\bibitem[\protect\citeauthoryear{{Sparre}, {Pfrommer}  \& {Ehlert}}{{Sparre}
  et~al.}{2020}]{SparrePfrommerEhlert2020}
{Sparre} M.,  {Pfrommer} C.,   {Ehlert} K.,  2020, \mn@doi [\mnras]
  {10.1093/mnras/staa3177}, \href
  {https://ui.adsabs.harvard.edu/abs/2020MNRAS.499.4261S} {499, 4261}

\bibitem[\protect\citeauthoryear{{Spilker} et~al.,}{{Spilker}
  et~al.}{2020a}]{SpilkerEtAl2020a}
{Spilker} J.~S.,  et~al., 2020a, \mn@doi [\apj] {10.3847/1538-4357/abc47f},
  \href {https://ui.adsabs.harvard.edu/abs/2020ApJ...905...85S} {905, 85}

\bibitem[\protect\citeauthoryear{{Spilker} et~al.,}{{Spilker}
  et~al.}{2020b}]{SpilkerEtAl2020b}
{Spilker} J.~S.,  et~al., 2020b, \mn@doi [\apj] {10.3847/1538-4357/abc4e6},
  \href {https://ui.adsabs.harvard.edu/abs/2020ApJ...905...86S} {905, 86}

\bibitem[\protect\citeauthoryear{{Spruit}}{{Spruit}}{2013}]{Spruit2013}
{Spruit} H.~C.,  2013, arXiv e-prints, \href
  {https://ui.adsabs.harvard.edu/abs/2013arXiv1301.5572S} {p. arXiv:1301.5572}

\bibitem[\protect\citeauthoryear{{Valdivia}, {Hennebelle}, {G{\'e}rin}  \&
  {Lesaffre}}{{Valdivia} et~al.}{2016}]{ValdiviaEtAl2016}
{Valdivia} V.,  {Hennebelle} P.,  {G{\'e}rin} M.,   {Lesaffre} P.,  2016,
  \mn@doi [\aap] {10.1051/0004-6361/201527325}, \href
  {http://adsabs.harvard.edu/abs/2016A%26A...587A..76V} {587, A76}

\bibitem[\protect\citeauthoryear{{Veilleux}, {Maiolino}, {Bolatto}  \&
  {Aalto}}{{Veilleux} et~al.}{2020}]{VeilleuxEtAl2020}
{Veilleux} S.,  {Maiolino} R.,  {Bolatto} A.~D.,   {Aalto} S.,  2020, \mn@doi
  [\aapr] {10.1007/s00159-019-0121-9}, \href
  {https://ui.adsabs.harvard.edu/abs/2020A&ARv..28....2V} {28, 2}

\bibitem[\protect\citeauthoryear{{Vieser} \& {Hensler}}{{Vieser} \&
  {Hensler}}{2007}]{VieserHensler2007}
{Vieser} W.,  {Hensler} G.,  2007, \mn@doi [\aap] {10.1051/0004-6361:20042120},
  \href {https://ui.adsabs.harvard.edu/abs/2007A&A...472..141V} {472, 141}

\bibitem[\protect\citeauthoryear{{Waagan}}{{Waagan}}{2009}]{Waagan2009}
{Waagan} K.,  2009, \mn@doi [Journal of Computational Physics]
  {10.1016/j.jcp.2009.08.020}, \href
  {http://adsabs.harvard.edu/abs/2009JCoPh.228.8609W} {228, 8609}

\bibitem[\protect\citeauthoryear{{Waagan}, {Federrath}  \&
  {Klingenberg}}{{Waagan} et~al.}{2011}]{Waagan2011}
{Waagan} K.,  {Federrath} C.,   {Klingenberg} C.,  2011, \mn@doi [Journal of
  Computational Physics] {10.1016/j.jcp.2011.01.026}, \href
  {http://adsabs.harvard.edu/abs/2011JCoPh.230.3331W} {230, 3331}

\bibitem[\protect\citeauthoryear{{Walch} et~al.,}{{Walch}
  et~al.}{2015}]{WalchEtAl2015}
{Walch} S.,  et~al., 2015, \mn@doi [\mnras] {10.1093/mnras/stv1975}, \href
  {http://adsabs.harvard.edu/abs/2015MNRAS.454..238W} {454, 238}

\bibitem[\protect\citeauthoryear{{Walter}, {Weiss}  \& {Scoville}}{{Walter}
  et~al.}{2002}]{WalterWeissScoville2002}
{Walter} F.,  {Weiss} A.,   {Scoville} N.,  2002, \mn@doi [\apjl]
  {10.1086/345287}, \href
  {https://ui.adsabs.harvard.edu/abs/2002ApJ...580L..21W} {580, L21}

\bibitem[\protect\citeauthoryear{{Wolfire}, {Hollenbach}, {McKee}, {Tielens}
  \& {Bakes}}{{Wolfire} et~al.}{1995}]{WolfireEtAl1995}
{Wolfire} M.~G.,  {Hollenbach} D.,  {McKee} C.~F.,  {Tielens} A.~G.~G.~M.,
  {Bakes} E.~L.~O.,  1995, \mn@doi [\apj] {10.1086/175510}, \href
  {http://adsabs.harvard.edu/abs/1995ApJ...443..152W} {443, 152}

\bibitem[\protect\citeauthoryear{{Wolfire}, {McKee}, {Hollenbach}  \&
  {Tielens}}{{Wolfire} et~al.}{2003}]{WolfireEtAl2003}
{Wolfire} M.~G.,  {McKee} C.~F.,  {Hollenbach} D.,   {Tielens} A.~G.~G.~M.,
  2003, \mn@doi [\apj] {10.1086/368016}, \href
  {http://adsabs.harvard.edu/abs/2003ApJ...587..278W} {587, 278}

\bibitem[\protect\citeauthoryear{{W{\"u}nsch}, {Walch}, {Dinnbier}  \&
  {Whitworth}}{{W{\"u}nsch} et~al.}{2018}]{WuenschEtAl2018}
{W{\"u}nsch} R.,  {Walch} S.,  {Dinnbier} F.,   {Whitworth} A.,  2018, \mn@doi
  [\mnras] {10.1093/mnras/sty015}, \href
  {http://esoads.eso.org/abs/2018MNRAS.475.3393W} {475, 3393}

\bibitem[\protect\citeauthoryear{{Xu} \& {Stone}}{{Xu} \&
  {Stone}}{1995}]{XuStone1995}
{Xu} J.,  {Stone} J.~M.,  1995, \mn@doi [\apj] {10.1086/176475}, \href
  {https://ui.adsabs.harvard.edu/abs/1995ApJ...454..172X} {454, 172}

\bibitem[\protect\citeauthoryear{{Zhou}}{{Zhou}}{2017}]{Zhou2017a}
{Zhou} Y.,  2017, \mn@doi [\physrep] {10.1016/j.physrep.2017.07.005}, \href
  {https://ui.adsabs.harvard.edu/abs/2017PhR...720....1Z} {720, 1}

\makeatother
\end{thebibliography}

\appendix
\section{Clouds in isolation}
\label{sec:nowind}

\begin{figure*}
\begin{minipage}{\textwidth}
\centering
\includegraphics[width=\textwidth]{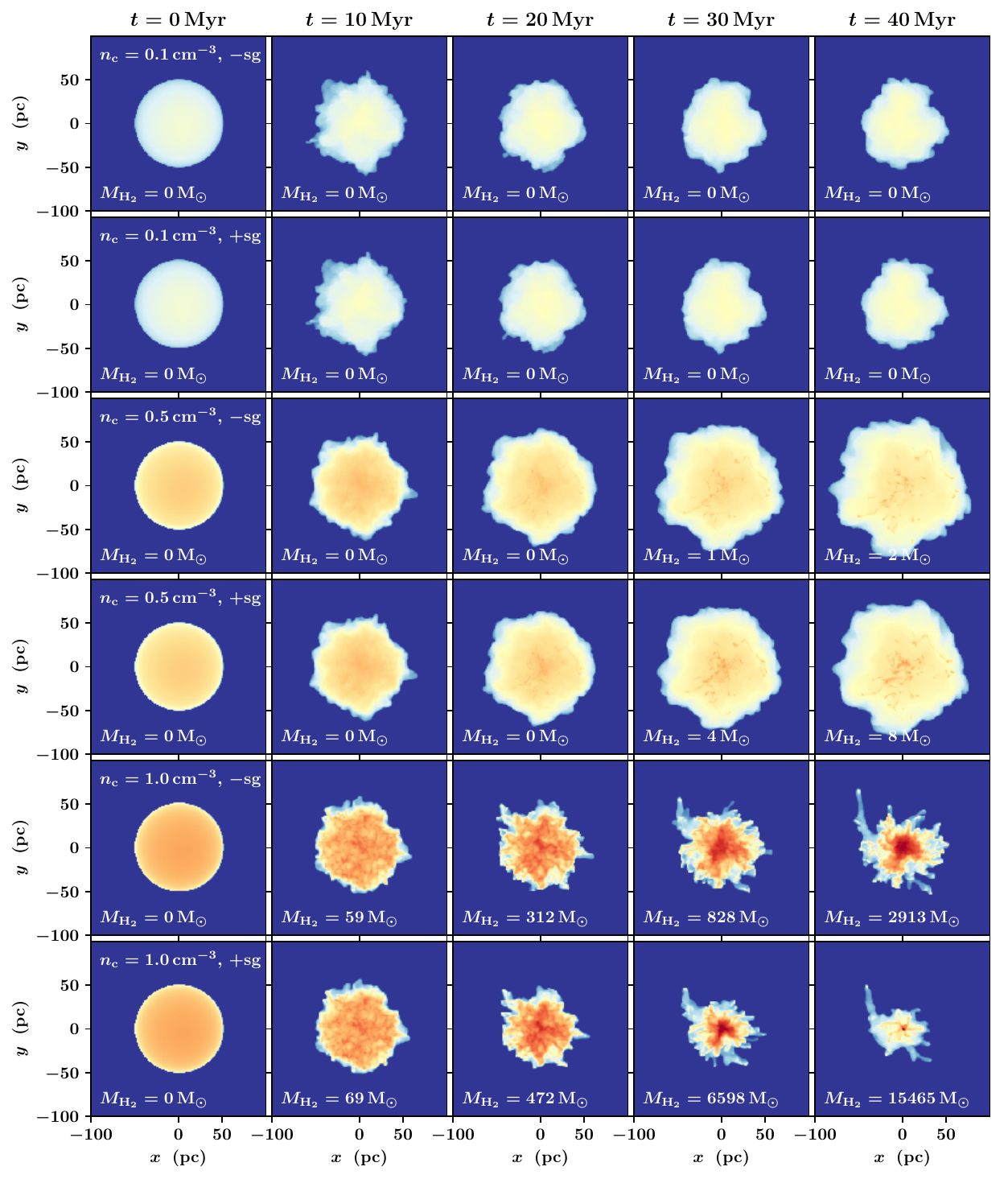}
\caption{Time evolution of the column density of all cloud simulations without the impact of a wind. From top to bottom we show the different simulations. From left to right the panels depict the column density at different times. In the panels we indicate the mass in molecular hydrogen.}
\label{fig:coldens-noW}
\end{minipage}
\end{figure*}

We also test the clouds in an isolated environment, i.e. without the impact of a wind. We simply evolve the clouds in the presence of background heating and cooling including the small turbulent motions inside the cloud but without the wind. The time evolution is depicted in the column density plots in Fig.~\ref{fig:coldens-noW}. From top to bottom we show the different simulation setups, from left to right we indicate the evolution over time. The first two rows with a cloud density of $n_\mathrm{c}=0.1\,\mathrm{cm}^{-3}$ do not show a significant change in morphology. The initial turbulence transports material from the cloud into the ambient hot medium. For low density clouds, this transported envelope is at a very low density, and it can thus be heated and evaporated efficiently by the hot environment. Therefore the spatial extent of the cloud remains similar to the initial conditions. In the case of the medium density clouds with $n_\mathrm{c}=0.5\,\mathrm{cm}^{-3}$ (two central rows) the turbulent motions are the same. However, as the gas is denser, it cannot be heated as fast and efficiently as in the low density case, which reduces the fraction of evaporated gas. Therefore, the cloud expands and slightly grows in size. The run including self-gravity indicates a higher mass in molecular hydrogen within the barely noticeable denser condensations in the central part of the cloud. Overall, the mass of molecular hydrogen is relatively low in both cases compared to the total mass of the cloud. The two bottom rows depict the column densities for the clouds with $n_\mathrm{c}=1.0\,\mathrm{cm}^{-3}$. Both clouds are dense enough for cooling to be efficient over the simulated time scale. This leads to a reduction in thermal pressure compared to the ambient pressure and causes the clouds to contract even without self-gravity. If self-gravity is included, the cloud condenses to small $\sim\mathrm{pc}$ size clump after 40\,Myr. The fraction of molecular hydrogen to the total hydrogen mass is 18 per cent in the case without gravity and 04 per cent in the corresponding self-gravitating run.

\section{Position of the cloud}
\label{sec:cloud-position}
\begin{figure*}
\begin{minipage}{0.8\textwidth}
\centering
\includegraphics[width=\textwidth]{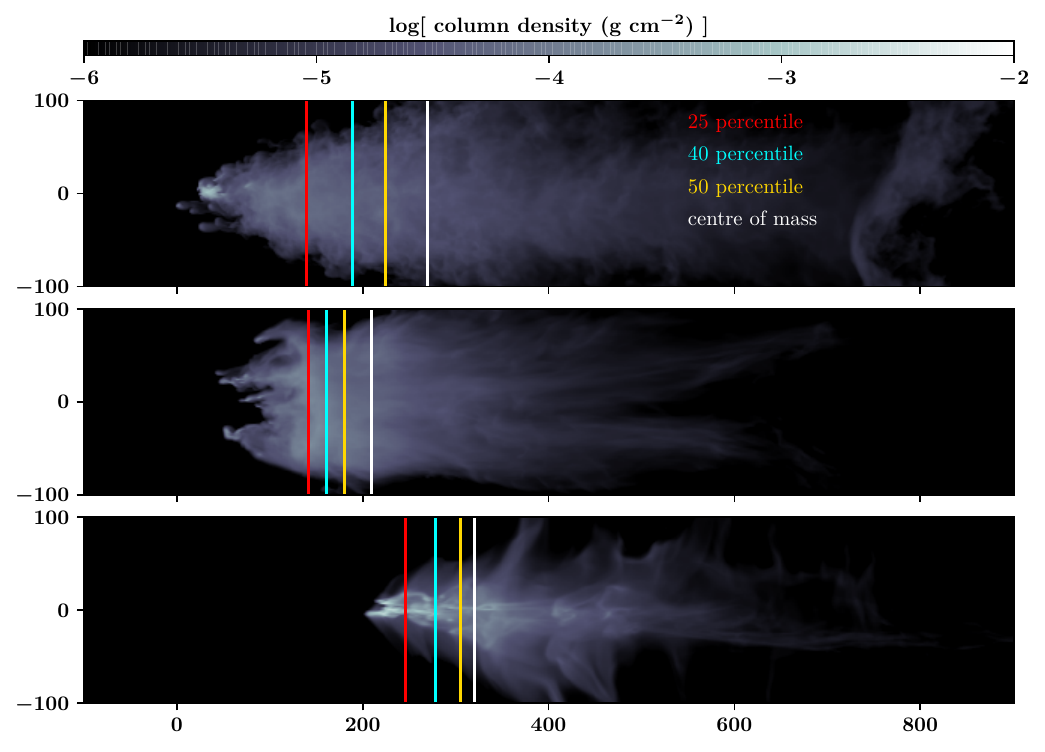}\\
\includegraphics[width=\textwidth]{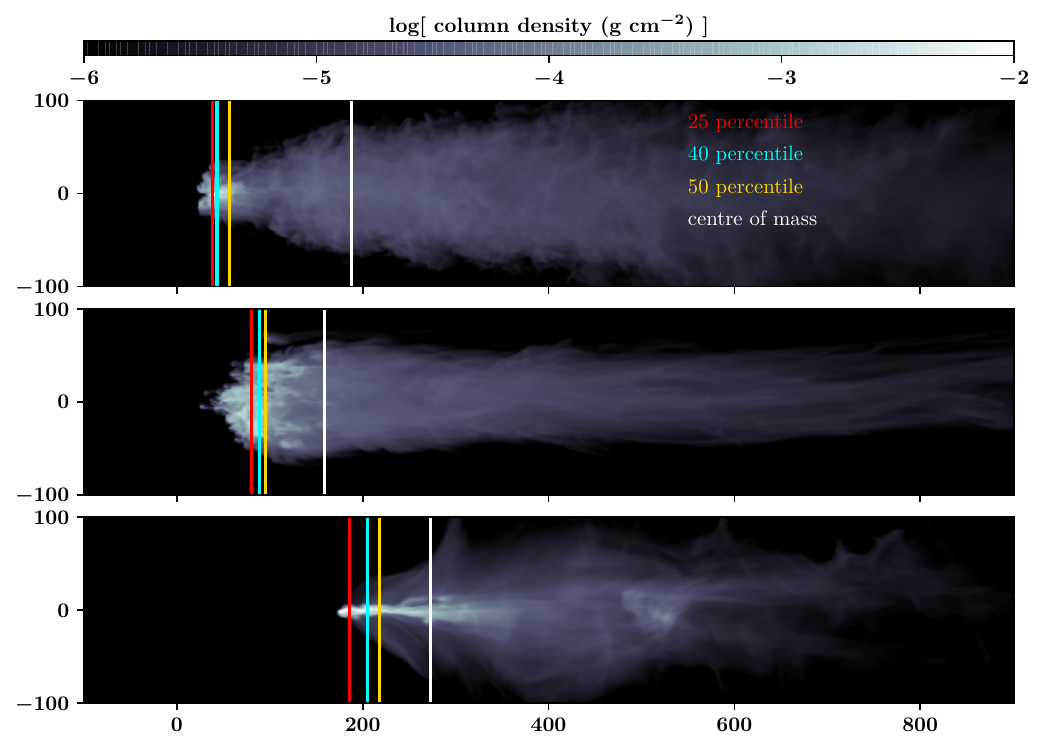}
\caption{Comparison of the position of the cloud over time. We visualized the 25, 40 and 50 percentile of the cumulative mass integrated along the $x$ direction (along the wind) as well as the centre of mass along $x$. We conclude that the 25 percentile is the best measure of the position of the cloud. The extended tail, which can contain a non-negligible mass of the initial cloud can shift the centre of mass and higher percentiles to locations aong $x$ that are not useful for measuring the position of the cloud.}
\label{fig:cloud-position}
\end{minipage}
\end{figure*}

It is not obvious what is the best definition of the position of the cloud as it evolves. In the initial conditions the centre of mass might be a simple measure to locate the cloud. If (significant) fractions of the mass are stripped during the time evolution and the front of the cloud that is exposed to the wind deforms, simple measures like the centre of mass or the first overdensity along the direction of the wind are not good indicators. We compare the following positions along the direction of the wind flow ($x$ direction):
\begin{itemize}
\item the centre of mass
\begin{equation}
x_\mathrm{c} = x_\mathrm{com} = \left(\sum_i m_i\right)^{-1}\,\sum_i x_i m_i,
\end{equation}
where $m_i$ is the mass of each individual cell, $x_i$ is the $x$ position and the sum includes all cells $i$.
\item the position of 25, 40, and 50 percentile of the cumulative mass integrated along the $x$ direction with the wind direction, i.e.
\begin{equation}
x_\mathrm{c} = x_k,\,\mathrm{where}\,\left(\sum_{ijk} m_{ijk}\right)^{-1}\, \sum_i \sum_{jk} m_{jik} = k,
\end{equation}
with $k=0.25$, $0.4$, and $0.6$, respectively, and the indices $i$, $j$, and $k$, correspond to the cell positions in $x$, $y$ and $z$ direction.
\end{itemize}
Fig.~\ref{fig:cloud-position} shows the four different measures for the cloud position for the clouds with $n_\mathrm{c}=0.5\,\mathrm{cm^{-1}}$ at $t=30\,\mathrm{Myr}$ (top figure) and for $n_\mathrm{c}=1\,\mathrm{cm^{-1}}$ at $t=40\,\mathrm{Myr}$ (bottom figure). The three panels in each figure depict from top to bottom the case of a non-magnetic wind, a wind with parallel field and a wind with perpendicular field. From the positions indicated with coloured vertical lines we conclude that the 25 percentile corresponds best with the actual cloud position.

\bsp	
\label{lastpage}
\end{document}